\documentclass[twocolumn,showpacs,aps,floatfix,prd,nofootinbib]{revtex4}
\usepackage{graphicx}
\usepackage{dcolumn}
\usepackage{amssymb}
\usepackage{amsmath}
\usepackage{colordvi}
\usepackage{color}
\usepackage{hhline}

\RequirePackage{xspace}
\newcommand{\BABARPubYear}    {10}
\newcommand{\BABARPubNumber}  {028}
\newcommand{\SLACPubNumber} {14317}

\usepackage{relsize}
\def\babar{\mbox{\slshape B\kern-0.1em{\smaller A}\kern-0.1em
    B\kern-0.1em{\smaller A\kern-0.2em R}}}

\def\Y#1S{\ensuremath{\Upsilon{(#1S)}}\xspace}

\def\pep2{PEP-II}

\long\def\inst#1{\par\nobreak\kern 4pt\nobreak
  {\it #1}\par\vskip 10pt plus 3pt minus 3pt}
  
\begin{document}

\begin{flushleft}
SLAC-PUB-\SLACPubNumber \\
\babar-PUB-\BABARPubYear/\BABARPubNumber \\
\end{flushleft}

\title{
Measurement of the $\gamma\gamma^\ast \to \eta$ and
$\gamma\gamma^\ast \to \eta^\prime$
transition form factors}

\author{P.~del~Amo~Sanchez}
\author{J.~P.~Lees}
\author{V.~Poireau}
\author{E.~Prencipe}
\author{V.~Tisserand}
\affiliation{Laboratoire d'Annecy-le-Vieux de Physique des Particules (LAPP), Universit\'e de Savoie, CNRS/IN2P3,  F-74941 Annecy-Le-Vieux, France}
\author{J.~Garra~Tico}
\author{E.~Grauges}
\affiliation{Universitat de Barcelona, Facultat de Fisica, Departament ECM, E-08028 Barcelona, Spain }
\author{M.~Martinelli$^{ab}$}
\author{D.~A.~Milanes}
\author{A.~Palano$^{ab}$ }
\author{M.~Pappagallo$^{ab}$ }
\affiliation{INFN Sezione di Bari$^{a}$; Dipartimento di Fisica, Universit\`a di Bari$^{b}$, I-70126 Bari, Italy }
% ...
\collaboration{The \babar\ Collaboration}
\noaffiliation

\begin{abstract}
We study the reactions $e^+e^-\to e^+e^-\eta^{(\prime)}$
in the single-tag mode and  measure the 
$\gamma\gamma^\ast \to \eta^{(\prime)}$ transition form factors
in the momentum-transfer range from 4 to 40 GeV$^2$. The analysis is
based on 469 fb$^{-1}$ of integrated luminosity collected at \pep2\ with
the \babar\ detector at $e^+e^-$ center-of-mass energies near 10.6 GeV.
\end{abstract}


\maketitle
\setcounter{footnote}{0}

\section{Introduction}\label{intro}
\begin{figure}
\includegraphics[width=.33\textwidth]{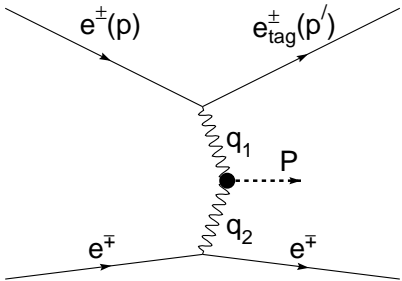}
\caption{The diagram for the $e^+e^-\to e^+e^-P$ two-photon
production process, where $P$ is a pseudoscalar meson.
\label{fig1}}
\end{figure}

In this article we report results from studies of the 
$\gamma^\ast\gamma \to P$ transition form factors, where
$P$ is a pseudoscalar meson. In our previous works~\cite{pi0ff,etacff}
the two-photon-fusion reaction  
\begin{equation}
e^+e^-\to e^+e^-P,
\end{equation}
illustrated by Fig.~\ref{fig1}, was used to measure the
$\pi^0$ and $\eta_c$ transition form factors. Here this technique is applied
to study the $\eta$ and $\eta^\prime$ form factors.
The transition form factor describes the effect of the strong interaction on
the $\gamma^\ast\gamma^\ast \to P$ transition. It is a function, $F(q^2_1,q^2_2)$, of 
the photon virtualities $q^2_i$.
We measure the differential cross sections for the processes 
$e^+e^-\to e^+e^-\eta^{(\prime)}$ in the single
tag mode where one of the outgoing electrons\footnote{Unless otherwise
specified, we use the term ``electron'' for either an electron or a positron.}
(tagged) is detected while the
other (untagged) is scattered at a small angle. The tagged electron 
emits a highly off-shell photon with the momentum transfer 
$q^2_1 \equiv -Q^2 = (p-p^\prime)^2$, where $p$ and $p^\prime$ are the four-momenta 
of the initial and final electrons. 
The momentum transfer to the untagged electron ($q^2_2$) is near zero. 
The form factor extracted from the single tag experiment is a function of one of 
the $q^2$'s: 
$F(Q^2)\equiv F(-Q^2,0)$.
To relate the differential cross section $d\sigma(e^+e^-\to e^+e^-P)/dQ^2$ to 
the transition form factor we use formulae equivalent to those
for the $e^+e^-\to e^+e^-\pi^0$ cross section in Eqs.~(2.1) and (4.5) of
Ref.~\cite{BKT}.

At large momentum transfer, perturbative QCD predicts that
the transition form factor can be represented as a convolution of a
calculable hard-scattering amplitude for $\gamma\gamma^\ast\to q\bar{q}$
with a nonperturbative meson distribution amplitude (DA) $\phi_{P}(x,Q^2)$~\cite{LB}.
The latter can be interpreted as the amplitude for the transition
of the meson with momentum $p_M$ into two quarks with momenta
$p_Mx$ and $p_M(1-x)$. The experimentally derived photon-meson transition
form factors can be used to test different models for the DA.

The $\eta$ and $\eta^\prime$ transition form factors have been measured 
in two-photon reactions in several previous experiments~\cite{etaff_0,etaff_1,etaff_2,etaff_3,CLEO}.
The most precise data for the $\eta^{(\prime)}$ at large $Q^2$ were obtained by the CLEO 
experiment~\cite{CLEO}. They cover the $Q^2$ region from 1.5 to about 20 GeV$^2$.
In this article we study the $\eta$ and $\eta^\prime$ form factors in the
$Q^2$ range from 4 to 40 GeV$^2$.

\section{The \babar\ detector and data samples}                 
\label{detector}                                                           
We analyze a data sample corresponding to an integrated
luminosity of about 469~fb$^{-1}$ recorded with                                                
the \babar\ detector~\cite{babar-nim} at the \pep2\                   
asymmetric-energy storage rings at the SLAC National Accelerator Laboratory.
At \pep2, 9-GeV electrons collide with     
3.1-GeV positrons to yield a center-of-mass (c.m.) energy near 10.58~GeV 
(i.e., the $\Upsilon$(4S) resonance peak). 
About 90\% of the data used in the present analysis were recorded on-resonance and 
about 10\% were recorded about 40 MeV below the resonance.

Charged-particle tracking is                                               
provided by a five-layer silicon vertex tracker  and                  
a 40-layer drift chamber, operating in a 1.5-T axial                 
magnetic field. The transverse momentum resolution                         
is 0.47\% at 1~GeV/$c$. Energies of photons and electrons                  
are measured with a CsI(Tl) electromagnetic calorimeter                    
with a resolution of 3\% at 1~GeV. Charged-particle                  
identification is provided by specific ionization (${\rm d}E/{\rm d}x$)            
measurements in the vertex tracker and drift chamber and by an internally reflecting           
ring-imaging Cherenkov detector. Electron identification
also makes use of the shower shape in the calorimeter and the ratio of shower energy to track
momentum. Muons are identified in the instrumented flux return
of the solenoid,
which consists of iron plates interleaved with either resistive plate chambers
or streamer tubes.

Signal $e^+e^-\to e^+e^-\eta^{(\prime)}$ and two-photon background processes are simulated
with the Monte Carlo (MC) event generator GGResRc~\cite{GGResRc}.  It uses the formula
for the differential cross section from Ref.~\cite{BKT} for pseudoscalar meson
production and the Budnev-Ginzburg-Meledin-Serbo formalism~\cite{Budnev} for the two-meson final states.
Because the $Q^2$ distribution is peaked near zero, the MC events
are generated with a restriction on the momentum transfer to one of the 
electrons: $Q^2 > 3$ GeV$^2$. This restriction corresponds to the
limit of detector acceptance for the tagged electron. The second electron is
required to have momentum transfer $-q_2^2 < 0.6$ GeV$^2$.
The experimental criteria providing these
restrictions for data events will be described in Sec.~\ref{evsel}.
The form factor is fixed to the constant value $F(0,0)$ in the simulation.

The GGResRc event generator includes next-to-leading-order radiative
corrections to the Born cross section calculated according to Ref.~\cite{RC}.
In particular, it generates extra soft photons emitted by
the initial- and final-state electrons.
The formulae from Ref.~\cite{RC} are modified to take into account
the hadron contribution to the vacuum polarization diagrams.
The maximum energy of the 
photon emitted from the initial state is restricted by the
requirement\footnote{Throughout this article an asterisk superscript
denotes quantities in the $e^+e^-$ c.m.\ frame. In this frame the positive
z-axis is defined to coincide with the $e^-$ beam direction.}
$E^\ast_\gamma < 0.05\sqrt{s}$, where $\sqrt{s}$ is the $e^+e^-$ c.m.\ energy. 
The generated events are subjected to a detailed
detector simulation based on GEANT4~\cite{GEANT4} 
and are reconstructed with the
software chain used for the experimental data. 
Temporal variations in                     
the detector performance                                                                  
and beam background conditions are taken into account.

\section{Event selection\label{evsel}}
The decay modes with two charged particles and two photons in the final
state,
$\eta^\prime \to \pi^+\pi^-\eta$, $\eta\to \gamma \gamma$ and
$\eta \to \pi^+\pi^-\pi^0$, $\pi^0 \to \gamma \gamma$, are used to reconstruct
$\eta^\prime$ and $\eta$ mesons, respectively. For the $e^+e^-\to e^+e^-\eta$
process, $\eta\to \pi^+\pi^-\pi^0$ is the only decay mode available for analysis at
BABAR. The trigger efficiency for events with $\eta$ decays
to $2\gamma$ and to $3\pi^0$ is very low.

Events  with at least three charged tracks and two photons are selected. 
Since a significant fraction
of signal events contains beam-generated spurious track and photon candidates,
one extra track and any number of extra photons are allowed in an event.
The tracks corresponding to the charged pions and electron
must have a point of closest approach to the nominal interaction
point (IP) that is within 2.5 cm along the beam axis and
less than 1.5 cm in the transverse plane. The track transverse 
momentum must be greater than 50 MeV/$c$.
 The identified pion candidates 
must have polar angles in the range $25.8^\circ<\theta<137.5^\circ$, while 
the track identified as an electron must be in the angular range 
$22.2^\circ<\theta<137.5^\circ$ (36.7--154.1$^\circ$ in the $e^+e^-$ 
c.m.\ frame). The angular
requirements are needed for good electron and pion identification. 
Electrons and pions are selected using a likelihood based identification
algorithm, which combines the measurements of the tracking system,
the Cherenkov detector, and the electromagnetic calorimeter.
The electron identification efficiency is about 98--99\%, with
a pion-misidentification probability below 10\%.
The pions are identified with about 98\% efficiency and a
electron-misidentification rate of about 7\%. To                     
recover electron energy loss due to bremsstrahlung, both internal and
in the detector material before the drift chamber,
the energy of any calorimeter shower close to the electron direction 
(within 35 and 50 mrad for the polar and azimuthal angle, respectively)
is combined                  
with the measured energy of the electron track. The resulting c.m.\ energy 
of the electron candidate must be greater than 1 GeV.

The photon candidates are required to have laboratory energies greater than 
50 MeV.
For the $e^+e^- \to e^+e^- \eta^\prime$ selection, 
two photon candidates are combined to form an $\eta$ candidate.
Their invariant mass is required to be in the range 0.480--0.600 GeV/$c^2$.
To suppress combinatorial background from spurious photons,
the photon helicity angle is required to satisfy the condition
$|\cos{\theta_h}|<0.9$\footnote{
Spurious photons tend to have low energy, and therefore align
opposite to the $\eta$/$\pi^0$ candidate's boost direction, whereas true
$\eta$/$\pi^0$ meson decays into two photons have a flat $\cos{\theta_h}$
distribution.}.
The helicity angle $\theta_h$ is defined in the $\eta$ rest frame as the angle
between the decay photon momentum and direction of the boost from 
the laboratory frame.
Each candidate is then fit with an $\eta$-mass constraint
to improve the precision of its momentum measurement. An $\eta^\prime$
candidate is formed from a pair of oppositely-charged pion candidates and
an $\eta$ candidate. The $\eta^\prime$ invariant mass must be in 
the range 0.920--0.995 GeV/$c^2$. 
The $\eta^\prime$ candidate is also then fit with a mass constraint.

Similar selection criteria are used for  
$e^+e^- \to e^+e^-\eta$ candidates.
An $\eta$ candidate is formed from a pair of oppositely charged 
pion candidates and a $\pi^0$ candidate, which is a combination of two photons
with invariant mass between 0.115 and 0.150 GeV/$c^2$ and
the cosine of the photon helicity angle $|\cos{\theta_h}|<0.9$.
The mass of the $\eta$ candidate must be in the selection region 0.48--0.62 GeV/$c^2$. 
 \begin{figure}
\includegraphics[width=.4\textwidth]{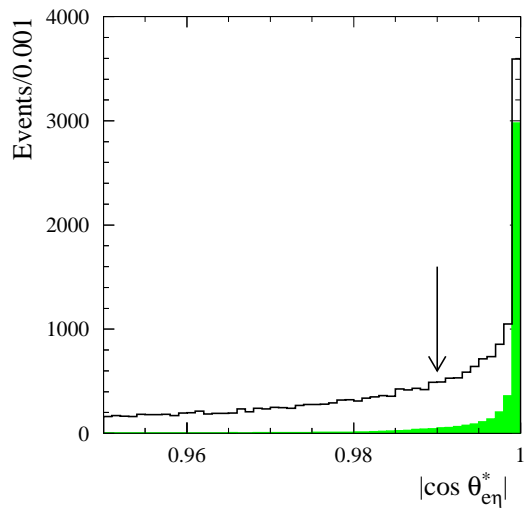}
\caption{The $|\cos{\theta^\ast_{e\eta}}|$
distribution for data events (solid histogram). The shaded histogram shows the same distributions 
for the $e^+e^- \to e^+e^-\eta$ simulation. Events with 
$|\cos{\theta^\ast_{e\eta}}|>0.99$ (indicated by the arrow) are retained.
\label{fig2}}
\end{figure}
\begin{figure}
\includegraphics[width=.4\textwidth]{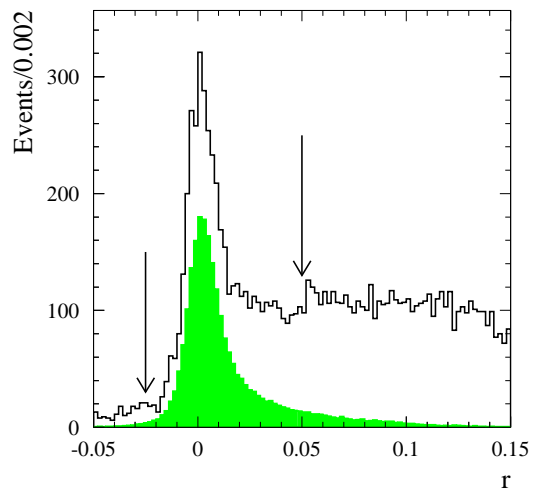}
\caption{The $r$ distributions 
for $e^+e^-\to e^+e^- \eta$ data (solid-line histogram) and 
signal simulation (shaded histogram). The arrows indicate the region used 
to select event candidates ($-0.025<r<0.05$).
\label{fig3}}
\end{figure}

Figure~\ref{fig2} shows the $|\cos{\theta^\ast_{e\eta}}|$
distribution for data and simulated $e^+e^- \to e^+e^-\eta$ events 
passing the selection criteria described above,
where $\theta^\ast_{e\eta}$ is the polar angle of the momentum vector of 
the $e\eta$ system in the $e^+e^-$ c.m.\ frame.
We require that $|\cos{\theta^\ast_{e\eta}}|$ be greater than 0.99. 
This condition effectively limits the value of the momentum transfer to the 
untagged electron ($q^2_2$) and
guarantees compliance with the condition $-q^2_2<0.6$ GeV$^2$ used in
the MC simulation.
The same condition $|\cos{\theta^\ast_{e\eta^\prime}}|>0.99$ is used to select
the $e^+e^- \to e^+e^-\eta^\prime$ event candidates.

The emission of extra photons by the electrons involved 
leads to a difference between
the measured and actual values of $Q^2$. In the case of initial-state 
radiation (ISR) $Q^2_{meas}=Q^2_{true}(1+r_\gamma)$, where 
$r_\gamma=2E^\ast_\gamma/\sqrt{s}$.
To restrict the energy of the ISR photon we use the parameter
\begin{equation}
r=\frac{{\sqrt{s}}-E^\ast_{e\eta^{(\prime)}}-|p^\ast_{e\eta^{(\prime)}}|}{\sqrt{s}},
\label{reqn}
\end{equation}
where $E^\ast_{e\eta^{(\prime)}}$ and $p^\ast_{e\eta^{(\prime)}}$ are the c.m.\ energy
and momentum of the detected $e\eta^{(\prime)}$ system. For ISR
this parameter coincides with $r_\gamma$ defined above. The $r$ distributions 
for data and simulated $e^+e^-\to e^+e^- \eta$ events
passing the selection criteria described above are
shown in Fig.~\ref{fig3}. For both processes under study, we select events 
with $-0.025<r<0.05$. It should be noted that this condition on $r$ ensures
compliance with the restriction $r_{\gamma}<0.1$ used in the simulation.
\begin{figure*}
\includegraphics[width=.4\textwidth]{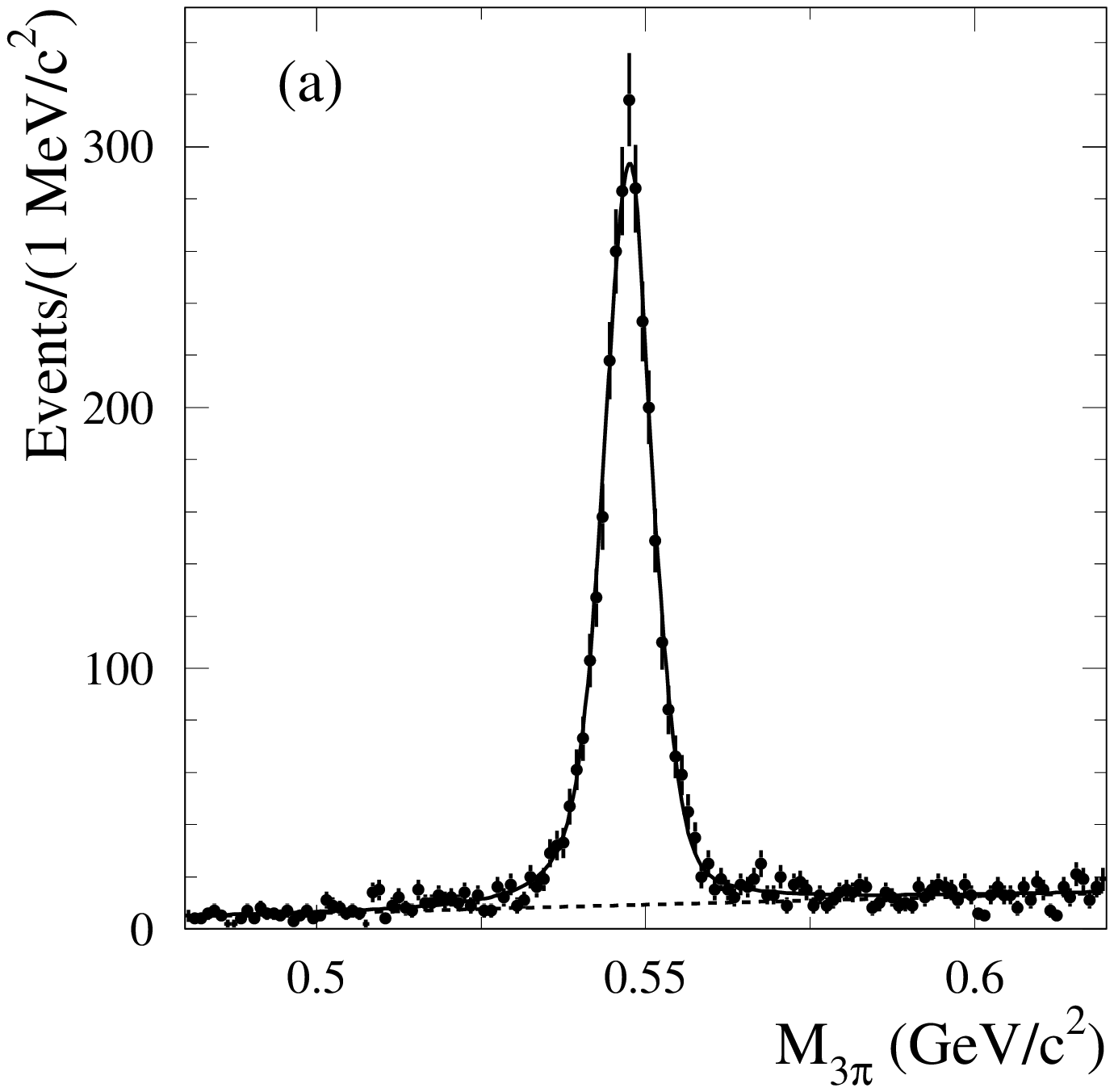}
\includegraphics[width=.4\textwidth]{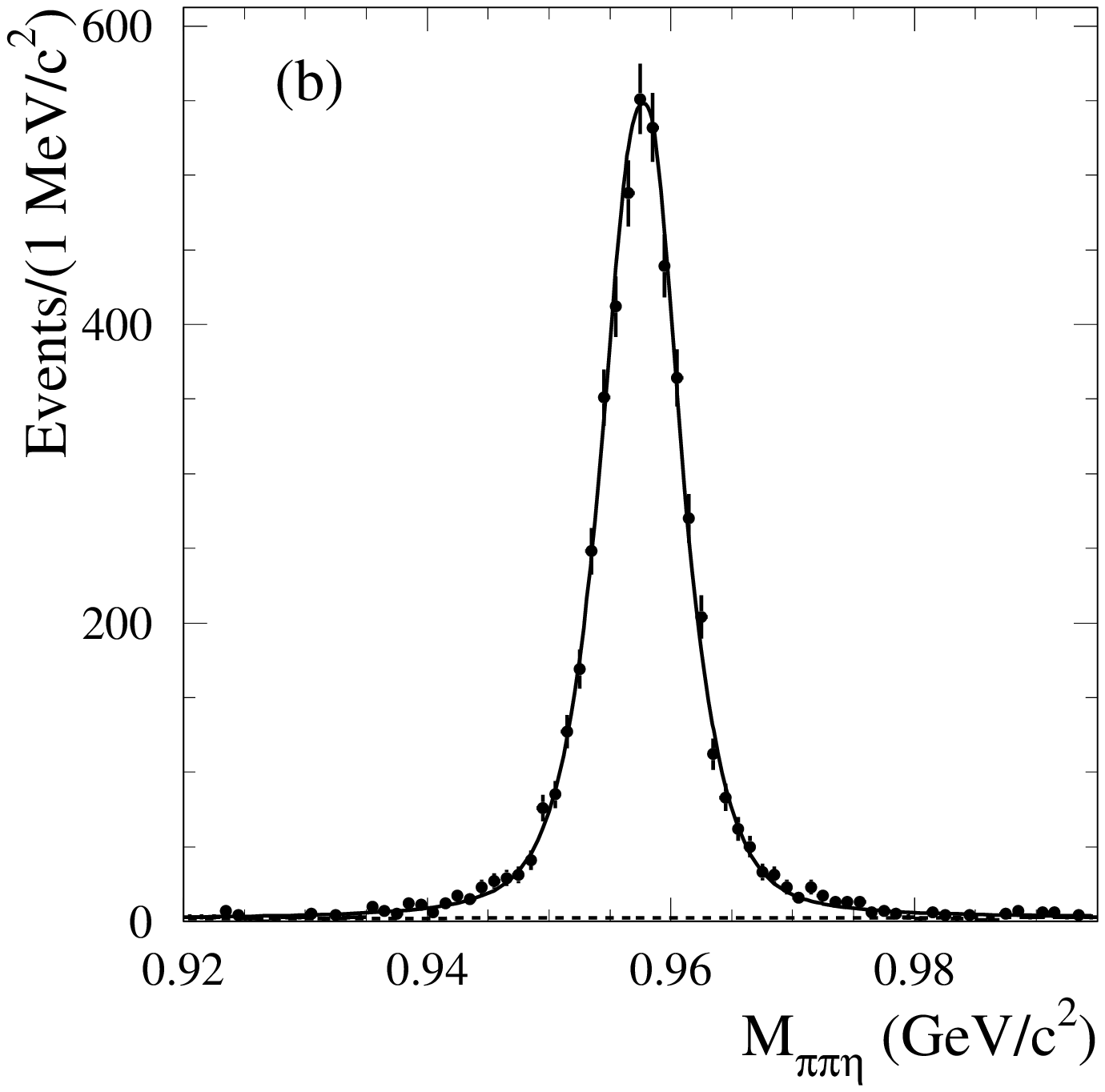}
\caption{
The (a) $\pi^+\pi^-\pi^0$ and (b) $\pi^+\pi^-\eta$ mass spectra for 
data events with $4 < Q^2  <40$ GeV$^2$. The solid curves are the results of the 
fits described 
in Sec.~\ref{fitting}. The dashed curves represent non-peaking background.
\label{fig4}}
\end{figure*}

For two-photon events with a tagged positron (electron),
the momentum of the detected $e\eta^{(\prime)}$ system in the $e^+e^-$ c.m.\ frame 
has a negative (positive) $z$-component, while events resulting 
from $e^+e^-$ annihilation are produced symmetrically. To suppress
the $e^+e^-$ annihilation background, event candidates 
with the wrong sign of the  momentum  $z$-component are removed.

The distributions of the invariant masses of $\eta$ and $\eta^\prime$ 
candidates for data events satisfying the 
selection criteria described above are shown in Fig.~\ref{fig4}.
For events with more than one $e^\pm\eta^{(\prime)}$ candidate
(about 5\% of the selected events), the candidate 
with smallest absolute value of the parameter $r$ is selected. 
Only events with $4 < Q^2  <40$ GeV$^2$ are included in the spectra of
Fig.~\ref{fig4}. For $Q^2 < 4$ GeV$^2$ the detection efficiency for 
single-tag two-photon $\eta$ and $\eta^\prime$ events is small 
(see Sec.~\ref{deteff}). In the region $Q^2 > 40$ GeV$^2$ we do not see 
evidence of $\eta$ or $\eta^\prime$ signal over background.
About 4350 and 5200 events survive the selection described above for $\eta$
and $\eta^\prime$, respectively.

\section{Fitting the $\pi^+\pi^-\pi^0$ and $\pi^+\pi^-\eta$ mass
spectra\label{fitting}}
\begin{figure*}
\includegraphics[width=.4\textwidth]{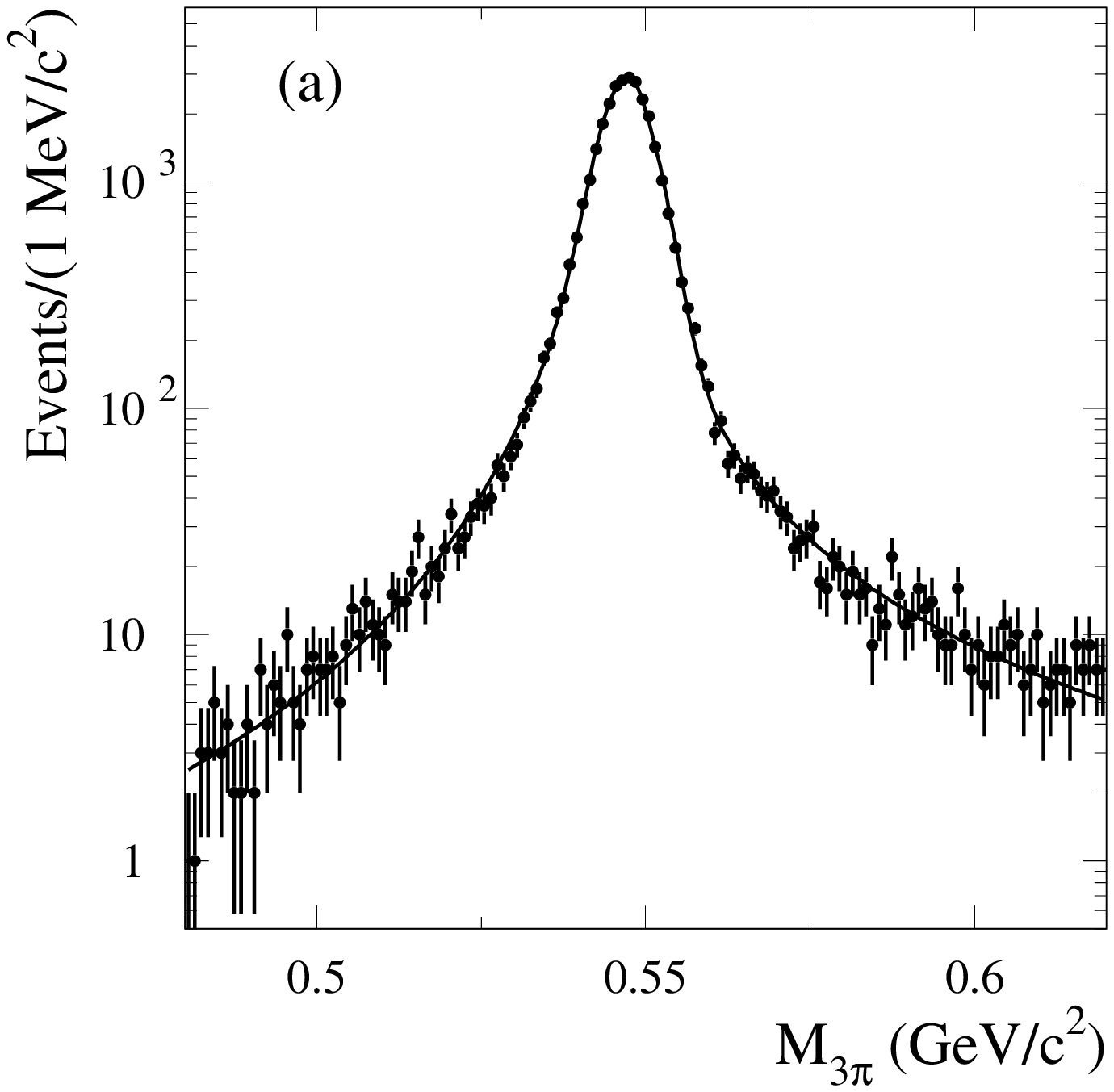}
\includegraphics[width=.4\textwidth]{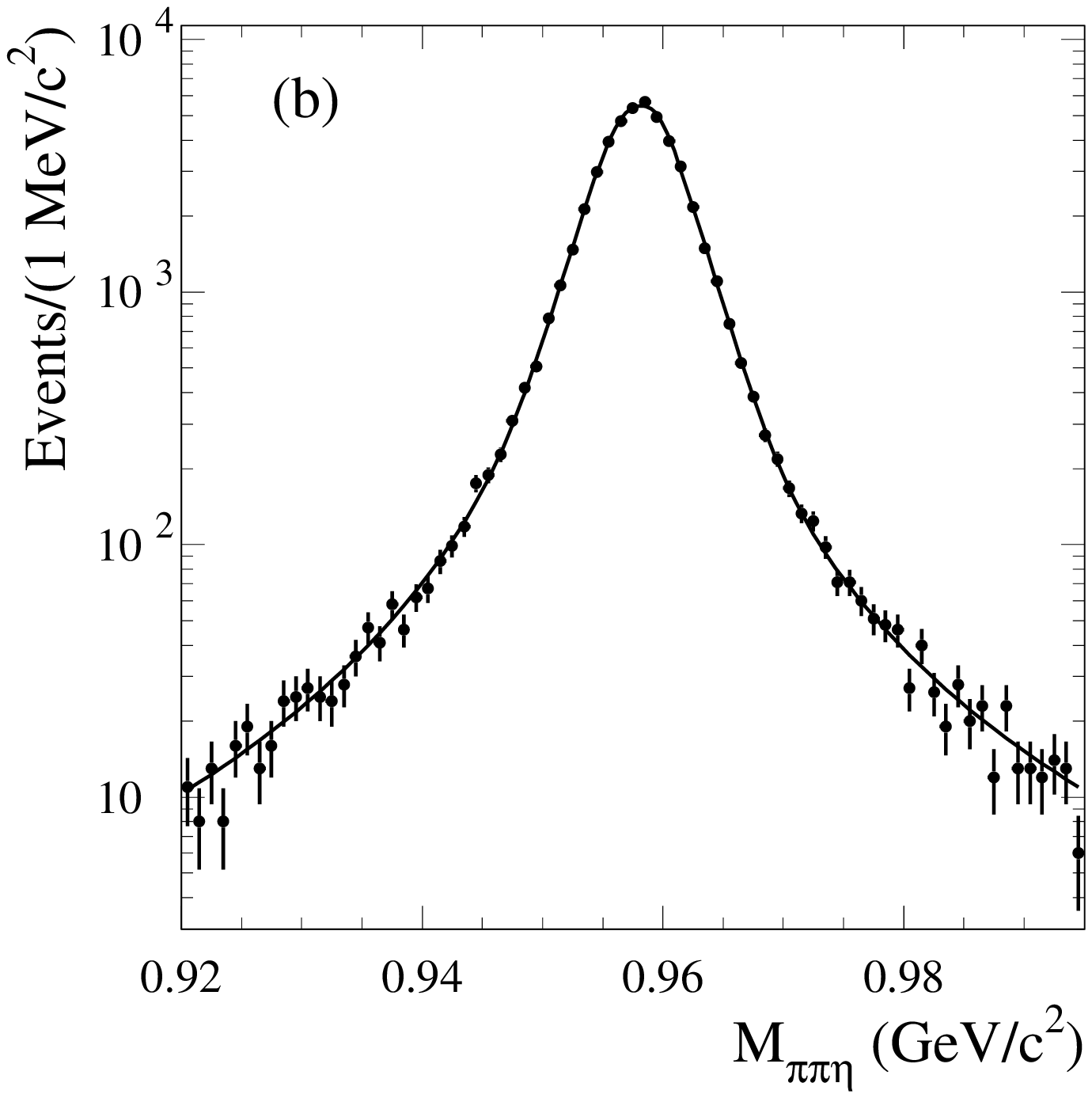}
\caption{
The $\pi^+\pi^-\pi^0$ and $\pi^+\pi^-\eta$  mass spectra for simulated 
(a) $e^+e^-\to e^+e^-\eta$ and (b)
$e^+e^-\to e^+e^-\eta^\prime$ events, respectively.
The curves represent the resolution functions described in the text.
\label{fig5}}
\end{figure*}
To determine the number of events containing an $\eta^{(\prime)}$,
we perform a binned likelihood fit to the spectra shown in
Fig.~\ref{fig4} with a sum of signal and background distributions.
The signal distributions are obtained by fitting mass spectra for simulated
signal events. The obtained functions then are modified to take into account
a possible difference between data and simulation in detector response. The
signal line shape in simulation is described by the following function:
\begin{equation}
F(x)=A[G(x)\sin^2{\zeta}+B(x)\cos^2{\zeta}],
\label{func}
\end{equation} 
where
\begin{equation}
G(x)=\exp \left (-\frac{(x-a)^2}{2\sigma^2} \right ),
\end{equation} 
\begin{equation}
B(x)=
\left\{ \begin{array}{ll}
\frac{(\Gamma_1/2)^{\beta_1}}{(a-x)^{\beta_1}+(\Gamma_1/2)^{\beta_1}} &
\mbox{if $x < a$};\\
\frac{(\Gamma_2/2)^{\beta_2}}{(x-a)^{\beta_2}+(\Gamma_2/2)^{\beta_2}} &
\mbox{if $x \geq a$},
\end{array}\right.
\end{equation} 
$\zeta$, $a$, $\sigma$, $\Gamma_1$, $\beta_1$, $\Gamma_2$, and $\beta_2$ 
are resolution function parameters, and $A$ is a normalization factor.
The $B(x)$ term is added to the Gaussian function to 
describe the asymmetric power-law tails of the detector resolution function.
The mass spectra for simulated signal events weighted
to yield the $Q^2$ dependencies observed in data and fitted curves are 
shown in Fig.~\ref{fig5}.
\begin{figure}
\includegraphics[width=.4\textwidth]{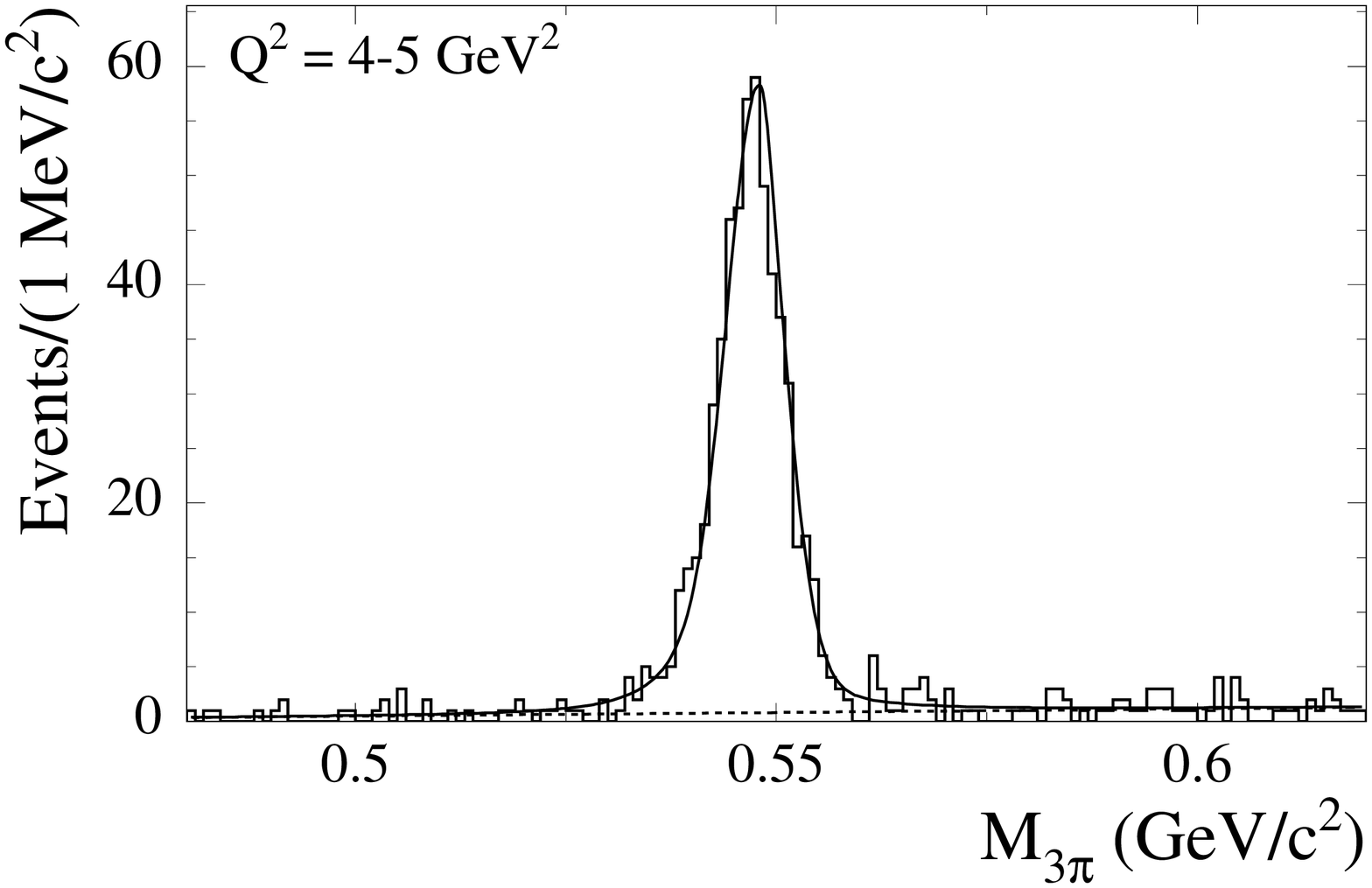}
\includegraphics[width=.4\textwidth]{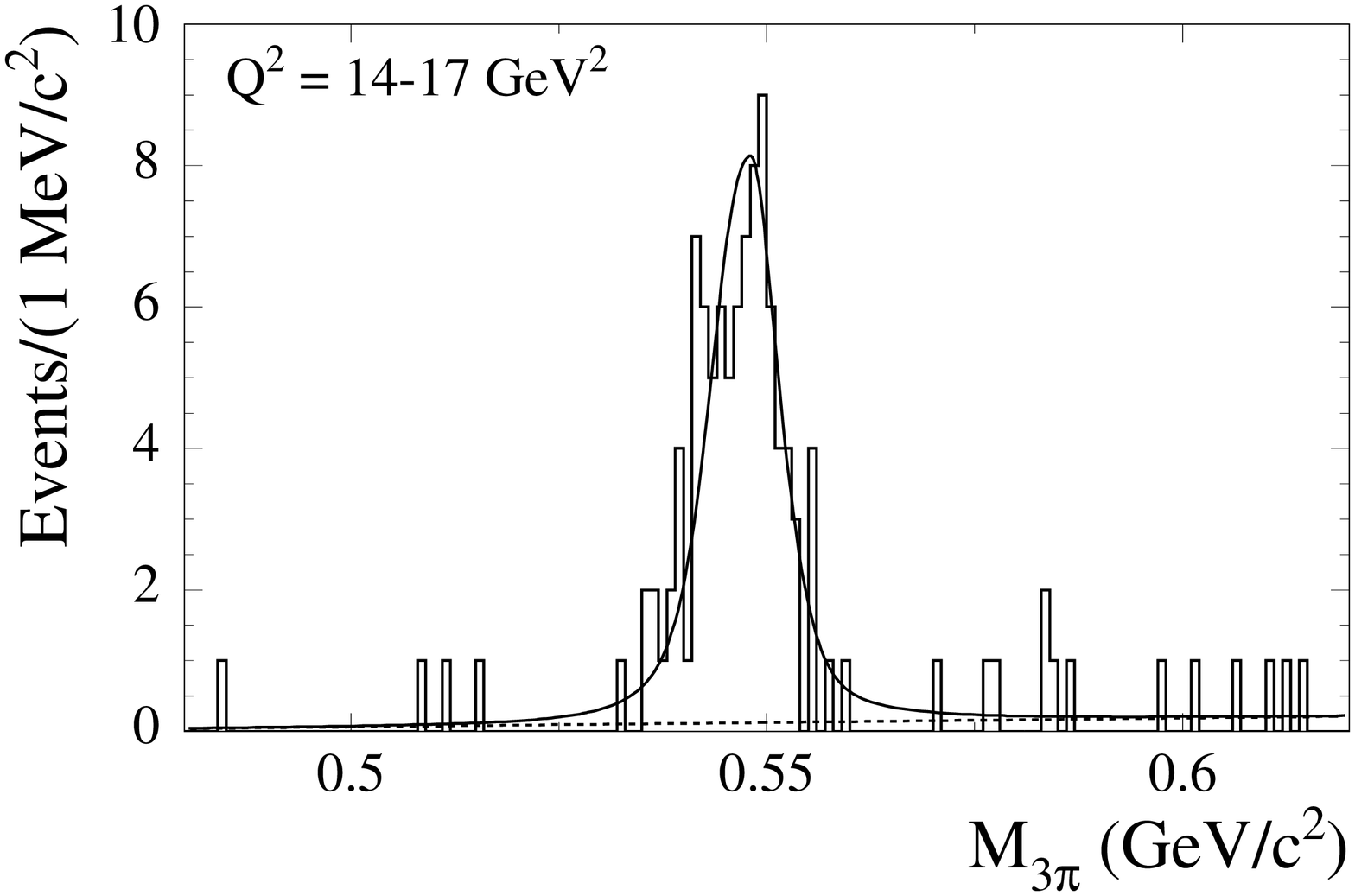}
\includegraphics[width=.4\textwidth]{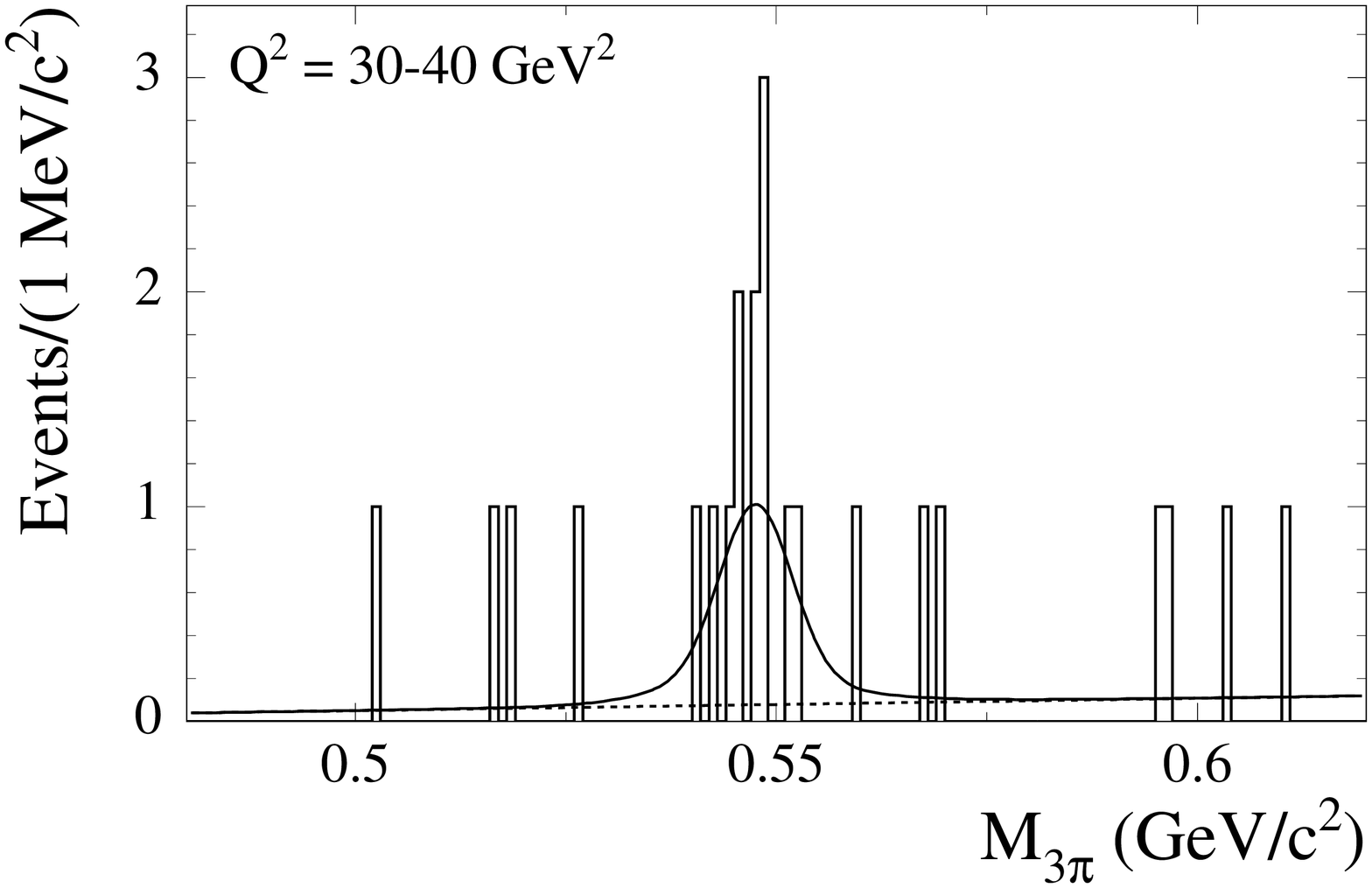}
\caption{The $\pi^+\pi^-\pi^0$ mass spectra for 
data events with $-0.025 < r < 0.025$ for
three representative $Q^2$ intervals. The solid
curves are the fit results. The dashed curves represent non-peaking background. 
\label{fig6}}
\end{figure}

When used in data, the parameters $\sigma$, $\Gamma_1$, $\Gamma_2$ and $a$
are modified to account for possible differences between data and simulation 
in resolution ($\Delta\sigma$) and mass scale calibration ($\Delta a$):
\begin{equation} 
\sigma^2=\left\{ \begin{array}{ll}
\sigma^2_{\rm MC}-\Delta\sigma^2 & \mbox{if $\Delta\sigma < 0$};\\
\sigma^2_{\rm MC}+\Delta\sigma^2 & \mbox{if $\Delta\sigma \geq 0$},
\end{array}\right.
\end{equation}
\begin{equation} 
\Gamma_i^2=\left\{ \begin{array}{ll}
\Gamma^2_{i,{\rm MC}}-(2.35\Delta\sigma)^2 & \mbox{if $\Delta\sigma < 0$};\\
\Gamma^2_{i,{\rm MC}}+(2.35\Delta\sigma)^2 & \mbox{if $\Delta\sigma \geq 0$},
\end{array}\right.
\end{equation}
\begin{equation}
a=a_{\rm MC}+\Delta a,
\end{equation}
where the subscript MC indicates the parameter value determined from the fit
to the simulated mass spectrum. The resolution and mass differences, $\Delta\sigma$
and $\Delta a$, are determined by a fit to data.

The background distribution is described by a linear function.
Five parameters are determined in the fit to the measured mass spectrum:
the number of 
$\eta^{(\prime)}$ events, $\Delta a$,
$\Delta\sigma$, and two background shape parameters.
The fitted curves are shown in Fig.~\ref{fig4}. The numbers of 
$\eta$ and $\eta^\prime$ events are found to be $3060\pm70$ and
$5010\pm90$, respectively. The mass shifts are $\Delta a=0.25\pm0.09$ MeV/$c^2$
for the $\eta$ and $\Delta a=-(0.48\pm0.06)$ MeV/$c^2$ for the $\eta^\prime$.
To check possible dependence of the mass shift on $Q^2$, separate 
fits are performed for two $Q^2$ regions: $4<Q^2<10$ GeV$^2$ and 
$10<Q^2<40$ GeV$^2$. The $\Delta a$ values obtained for these regions
agree with each other both for $\eta$ and $\eta^\prime$. In contrast,
the values of $\Delta\sigma$ are found to be strongly dependent on
$Q^2$, changing from $0.9\pm0.3$ MeV/$c^2$ for $4<Q^2<10$ GeV$^2$ to
$-(1.0\pm0.6)$ MeV/$c^2$ for $10<Q^2<40$ GeV$^2$. It should be noted that
the mass resolution for $\eta$ and $\eta^\prime$ is about 4 MeV/$c^2$.
The data-MC difference, $\Delta\sigma\sim 1$ MeV/$c^2$,  
corresponds to a small ($\sim 3$\%) change in the mass resolution when added in
quadrature.

\begin{table*}
\caption{The $Q^2$ interval, number of detected $e^+e^-\to e^+e^-\eta$ signal
events ($N_{s}$),
number of peaking-background events ($N_{b}$), efficiency correction 
($\delta_{\rm total}$), number of signal events corrected for data-MC 
difference and resolution effects ($N_{\rm corr}^{\rm unfolded}$), and
detection efficiency obtained from simulation ($\varepsilon$).
The first and second errors on $N_{s}$ and $N_{\rm corr}^{\rm unfolded}$ are 
statistical and systematic, respectively. The errors on $N_{b}$ are 
statistical and systematic combined in quadrature.
\label{tab10}}
\begin{ruledtabular}
\begin{tabular}{cccccc}
$Q^2$ interval (GeV$^2$) & $N_{s}$ & $N_{b}$ & $\delta_{\rm total} (\%)$ & $N_{\rm corr}^{\rm unfolded}$ & $\varepsilon (\%)$ \\
\hline 
  4--5  & $638\pm31\pm16   $ & $ 53\pm27   $ & $-1.4$ & $634\pm34\pm18$    &  6.3 \\ 
  5--6  & $625\pm34\pm19   $ & $ 89\pm34   $ & $-1.6$ & $641\pm38\pm22$    & 13.0 \\ 
  6--8  & $622\pm36\pm23   $ & $ 97\pm37   $ & $-1.7$ & $634\pm39\pm25$    & 14.7 \\ 
  8--10 & $349\pm26\pm12   $ & $ 43\pm23   $ & $-2.0$ & $359\pm29\pm14$    & 18.7 \\ 
 10--12 & $212\pm20\pm 7   $ & $ 15\pm16   $ & $-2.3$ & $224\pm22\pm8$     & 22.6 \\ 
 12--14 & $104\pm14\pm 4   $ & $ 13\pm11   $ & $-2.1$ & $105\pm17\pm5$     & 22.9 \\ 
 14--17 & $109\pm13\pm 3   $ & $0.0\pm 9.2 $ & $-2.0$ & $116\pm15\pm4$     & 22.2 \\ 
 17--20 & $40.5\pm8.3\pm1.2$ & $0.7\pm 5.6 $ & $-2.3$ & $41.2\pm9.5\pm1.4$ & 21.3 \\ 
 20--25 & $32.5\pm7.4\pm0.8$ & $0.0\pm 4.2 $ & $-2.4$ & $34.4\pm8.3\pm0.9$ & 19.6 \\ 
 25--30 & $13.7\pm5.3\pm0.5$ & $3.1\pm 3.5 $ & $-2.7$ & $14.2\pm6.0\pm0.6$ & 18.0 \\ 
 30--40 & $13.0\pm4.8\pm0.3$ & $0.5\pm 3.7 $ & $-2.7$ & $14.1\pm5.3\pm0.3$ & 15.7 \\
\end{tabular}
\end{ruledtabular} 
\end{table*}
A fitting procedure similar to that described above is applied 
in each of the eleven $Q^2$ intervals indicated in Table~\ref{tab10}. 
The parameters of the mass resolution function are taken from the fit to
the mass spectrum for simulated events in the corresponding $Q^2$ interval.
The $\eta$ and $\eta^\prime$ masses are fixed to the values obtained
from the fit to the spectra of Fig.~\ref{fig4}. The $\Delta\sigma$ parameter
is set to zero. Fits with $\Delta\sigma=0.9$ MeV/$c^2$ and  
$\Delta\sigma=-1.0$ MeV/$c^2$ are also performed.
The differences between the results of the fits with zero and non-zero 
$\Delta\sigma$ provide an estimate of the systematic uncertainty associated
with the data-MC simulation difference in the detector mass resolution.

For the analysis of the $e^+e^-\to e^+e^- \eta$ process, the numbers of events 
containing an $\eta$ are determined in two regions of the 
parameter $r$: $-0.025 < r < 0.025$ ($N_1$) and $0.025 < r < 0.050$ ($N_2$).  
The $N_1$ and $N_2$ values are used to determine the numbers of signal events ($N_s$)
and background events peaking at the $\eta$ mass ($N_b$) as described
in Sec.~\ref{background}. These values are listed in Table~\ref{tab10}. 
The $\pi^+\pi^-\pi^0$ mass 
spectra and fitted curves for three representative $Q^2$ intervals 
are shown in Fig.~\ref{fig6}. The spectra shown are obtained for the 
$-0.025 < r < 0.025$ regions; the  $0.025 < r < 0.050$ regions contain only 
10--13\% of the signal events and are used mainly to estimate backgrounds.

\begin{figure*}
\includegraphics[width=.32\textwidth]{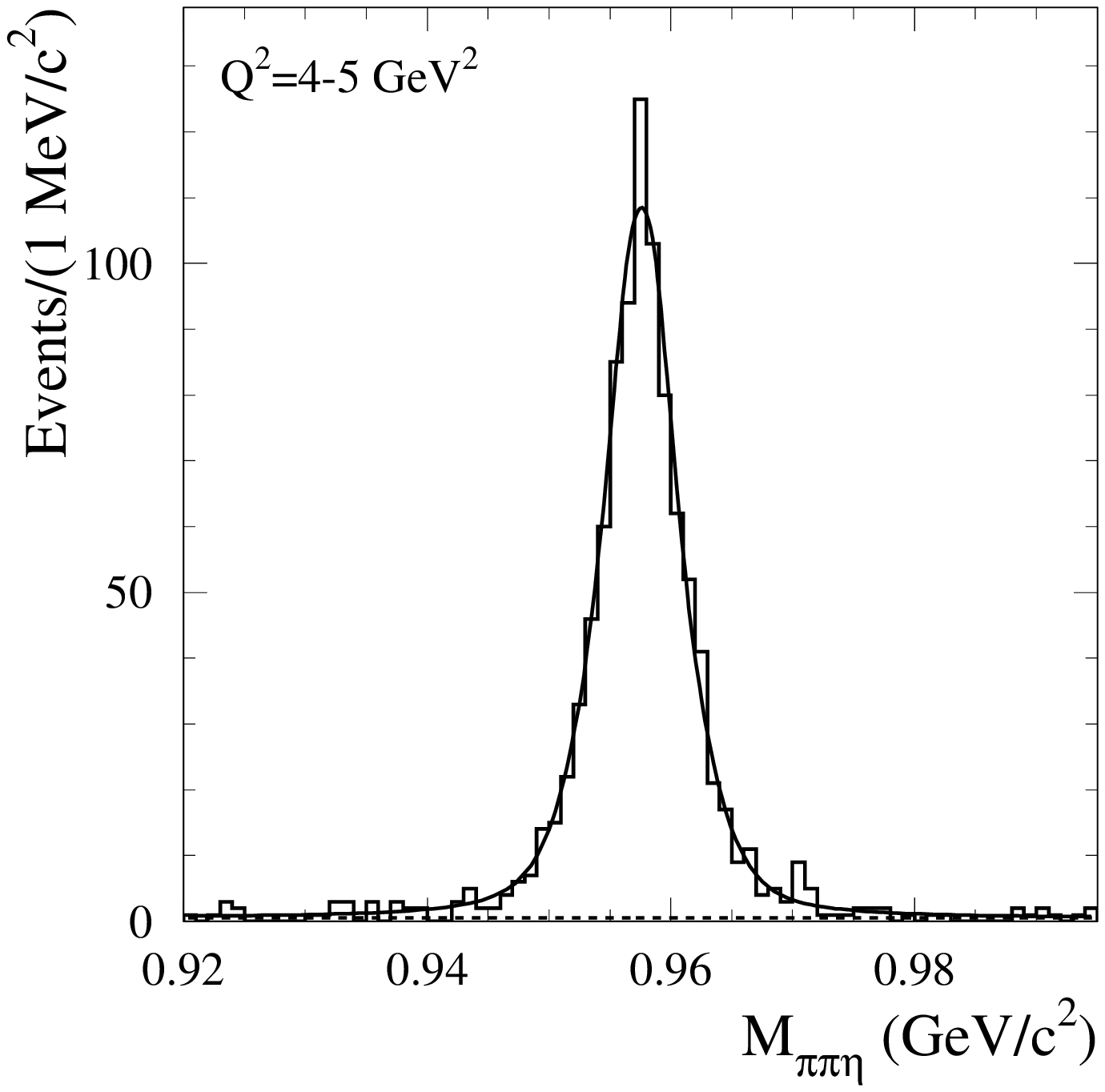}
\includegraphics[width=.32\textwidth]{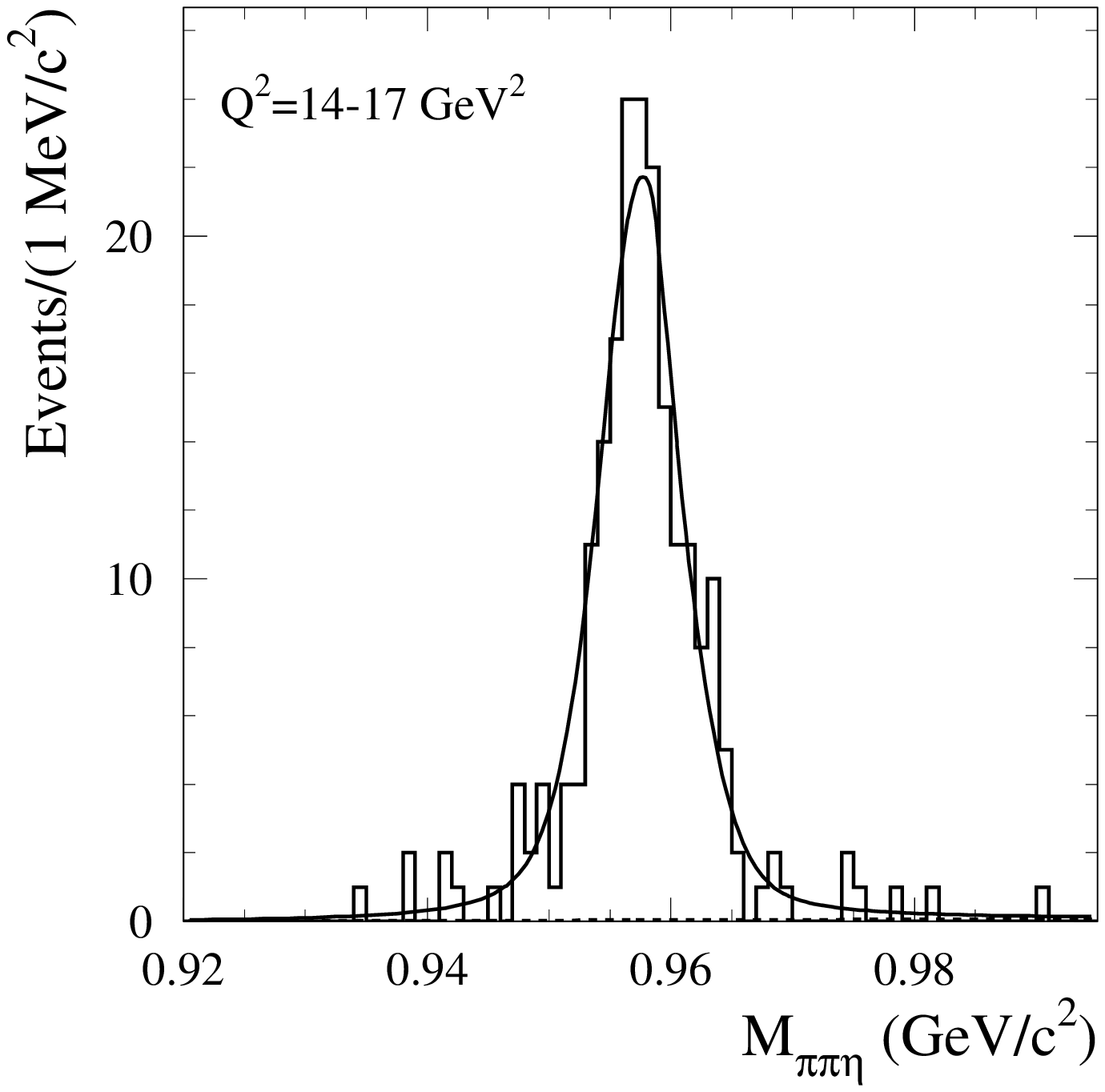}
\includegraphics[width=.32\textwidth]{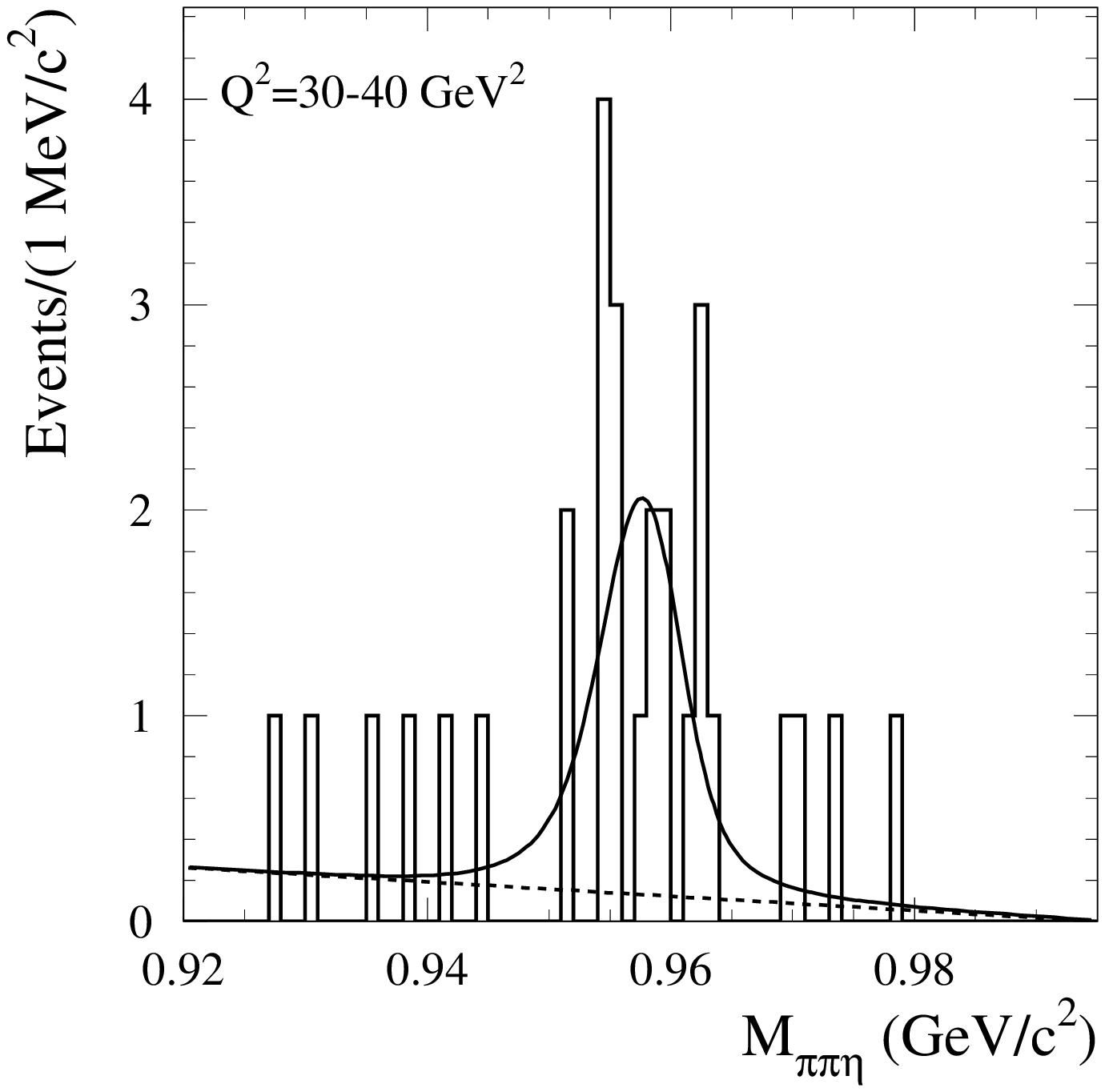}
\caption{The $\pi^+\pi^-\eta$ mass spectra for data events for
three representative $Q^2$ intervals. The solid curves are the fit results.
The dashed lines represent non-peaking background. 
\label{fig7}}
\end{figure*}

For the $e^+e^-\to e^+e^-\eta^\prime$ process, background is assumed to be
small. There is no need to separate events into two $r$ regions. 
The $\pi^+\pi^-\eta$ mass spectra and fitted curves for three representative
$Q^2$ intervals are shown in Fig.~\ref{fig7}. The numbers of
signal $\eta^\prime$ events obtained from the fits are listed in 
Table~\ref{tab11}.
\begin{table*}
\caption{The $Q^2$ interval, number of detected $\eta^\prime$ signal 
events ($N_{s}$), 
number of peaking-background events ($N_{b}$), efficiency correction 
($\delta_{\rm total}$), number of signal events corrected for data-MC 
difference and resolution effects ($N_{\rm corr}^{\rm unfolded}$), and
detection efficiency obtained from simulation ($\varepsilon$).
The first and second errors on $N_{s}$ and $N_{\rm corr}^{\rm unfolded}$ are 
statistical and systematic, respectively.
\label{tab11}}
\begin{ruledtabular}
\begin{tabular}{cccccc}
$Q^2$ interval (GeV$^2$) & $N_{s}$ & $N_{b}$ & $\delta_{\rm total} (\%)$ & $N_{\rm corr}^{\rm unfolded}$ & $\varepsilon (\%)$ \\
\hline 
  4--5  & $ 950\pm32\pm 5    $ & $0.0\pm 0.0$ & $ -0.4$ & $ 936\pm34\pm 6$ &  5.7 \\
  5--6  & $1013\pm33\pm 6    $ & $0.0\pm 0.0$ & $ -0.6$ & $1015\pm36\pm 7$ & 12.5 \\
  6--8  & $1185\pm36\pm 5    $ & $0.0\pm 0.0$ & $ -0.7$ & $1207\pm38\pm 6$ & 14.3 \\
  8--10 & $ 710\pm28\pm 3    $ & $0.0\pm 0.0$ & $ -1.0$ & $ 716\pm30\pm 4$ & 19.9 \\
 10--12 & $ 454\pm22\pm 4    $ & $0.0\pm 0.0$ & $ -1.2$ & $ 467\pm25\pm 4$ & 26.4 \\
 12--14 & $ 243\pm16\pm 1    $ & $0.0\pm 0.0$ & $ -1.0$ & $ 250\pm19\pm 1$ & 28.1 \\
 14--17 & $ 207\pm15\pm 2    $ & $0.0\pm 0.0$ & $ -0.8$ & $ 214\pm17\pm 2$ & 28.1 \\
 17--20 & $ 108\pm10\pm 1    $ & $0.0\pm 0.0$ & $ -0.8$ & $ 112\pm12\pm 1$ & 26.8 \\
 20--25 & $80.0\pm 9.0\pm 0.1$ & $0.0\pm 0.0$ & $ -1.0$ & $  82.5\pm 9.9\pm 0.2$ & 26.3 \\
 25--30 & $30.2\pm 5.9\pm 0.2$ & $1.0\pm 1.0$ & $ -1.3$ & $  31.7\pm 6.7\pm 0.2$ & 25.6 \\
 30--40 & $17.2\pm 5.4\pm 0.1$ & $2.0\pm 1.4$ & $ -1.4$ & $  18.1\pm 5.8\pm 0.1$ & 22.5 \\
\end{tabular}
\end{ruledtabular}
\end{table*}

\section{Peaking background estimation and subtraction}\label{background}
Background events containing true $\eta$ or $\eta^\prime$ mesons might arise from 
$e^+e^-$ annihilation, and two-photon processes with higher 
multiplicity final states than our signal events. The $e^+e^-$ annihilation 
background is studied in Sec.~\ref{background1}. In 
Sec.~\ref{background2} we use events with an extra $\pi^0$ to estimate 
the level of the two-photon background and study its characteristics. In 
Sec.~\ref{background3} we develop a method of background subtraction 
based on the difference in the $r$ distributions for signal and background
events. This method gives an improvement in accuracy compared to the previous 
one described in Sec.~\ref{background2} 
and has a lower sensitivity to the model used for background simulation. 

\subsection{$e^+e^-$ annihilation background}\label{background1}
The background from $e^+e^-$ annihilation can
be estimated using events with the wrong sign of the $e^\pm\eta^{(\prime)}$
momentum  $z$-component. The numbers of background events from $e^+e^-$ 
annihilation in the wrong- and right-sign data samples are expected to be 
approximately the same, but their $Q^2$ distributions are quite different.
The $Q^2$ distribution expected for right-sign background events coincides
with the $Q^2_{ws}$ distribution for wrong-sign events, where
$Q^2_{ws}$ is the squared difference
between the four-momenta of the detected positron (electron) and the initial 
electron (positron). 

In the $Q^2_{ws}$ region from 4 to 40 GeV$^2$ we observe 3 wrong-sign events 
in the $\eta^\prime$ data sample, all peaking at the $\eta^\prime$ mass, 
and 9 events in the $\eta$ data sample, 5 of which are in the 0.530--0.565 
GeV/$c^2$ mass window. The contribution from non-$\eta$ events to this mass 
window is estimated to be 0.3 events. A possible source of 
these events is the $e^+e^-\to X \gamma$ process, where $X$ is a hadronic
system containing an $\eta$ or $\eta^\prime$ meson, for example, 
$\pi^+\pi^-\eta^\prime$, with the photon emitted along the beam axis.

The $Q^2_{ws}$ distribution for the wrong-sign events is used to estimate
the $Q^2$ distribution for $e^+e^-$ annihilation background in 
the right-sign data sample. 
The fraction of $e^+e^-$ annihilation events in the $\eta^{(\prime)}$ 
data sample is about $10^{-3}$. However, such events are the main
contribution to the peaking background in high $Q^2$ bins and cannot 
be neglected.
For the $e^+e^-\to e^+e^-\eta^\prime$ process, for which 
we do not observe a significant two-photon background
(see Sec.\ref{background2}), the 3 background events from $e^+e^-$ annihilation
are subtracted from the two highest $Q^2$ intervals (see Table~\ref{tab11}).

For the $e^+e^-\to e^+e^-\eta$ process, the $e^+e^-$ annihilation events
are effectively subtracted with the procedure developed for subtraction
of two-photon background (see Sec.\ref{background3}). The procedure exploits
the difference between the $r$ distributions for signal and background events.
The $r$ distribution for the $e^+e^-$ annihilation events (3 of 5 events 
have $r>0.025$) is close to that for two-photon background.

In future high statistics measurements of the meson-photon form factors
at Super $B$ factories $e^+e^-$ annihilation will be the  dominant 
background in the high $Q^2$ region ($Q^2\gtrsim 50$ GeV$^2$).

\subsection{Two-photon background}\label{background2}
\begin{figure*}
\includegraphics[width=.4\textwidth]{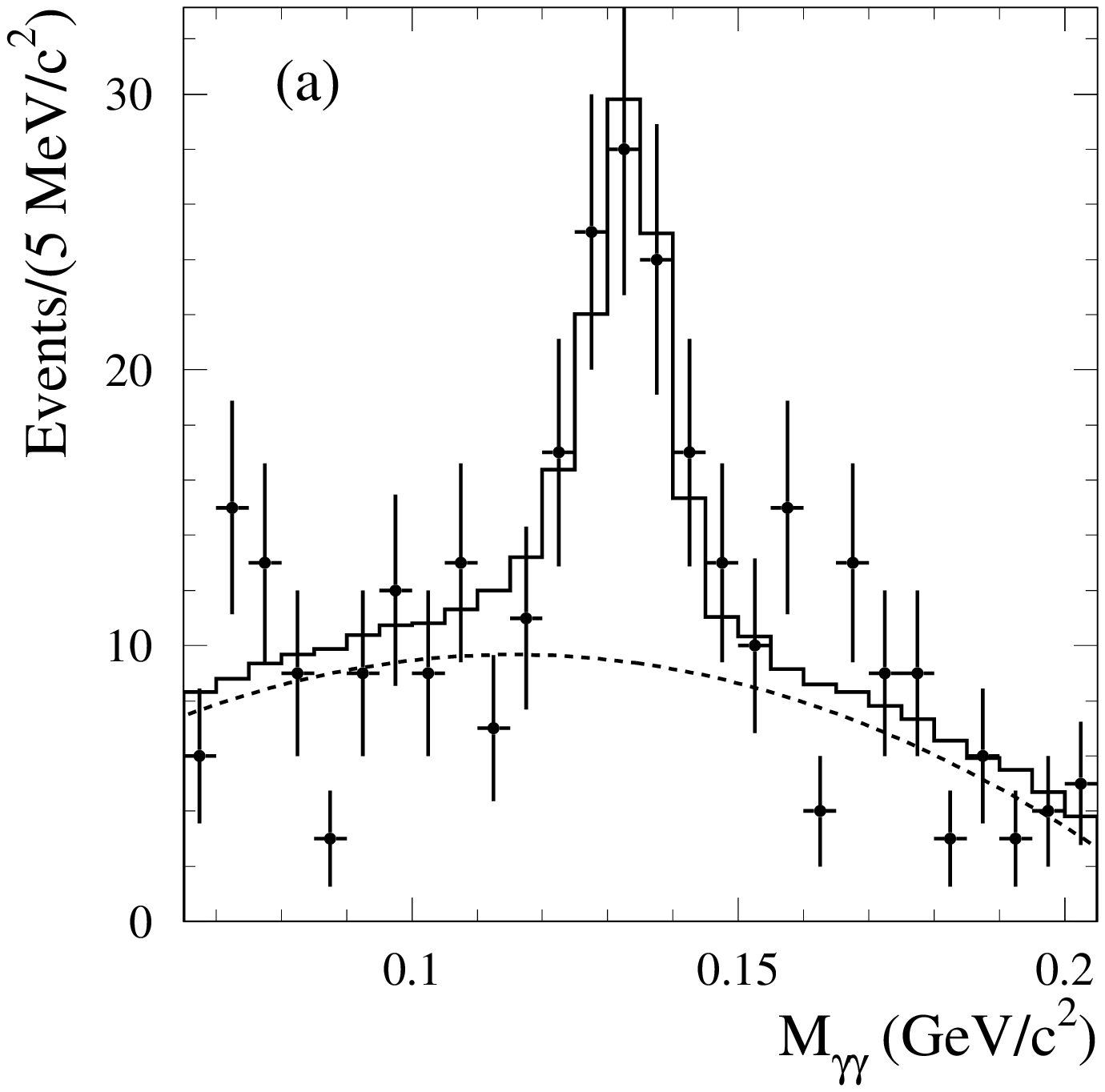}
\includegraphics[width=.4\textwidth]{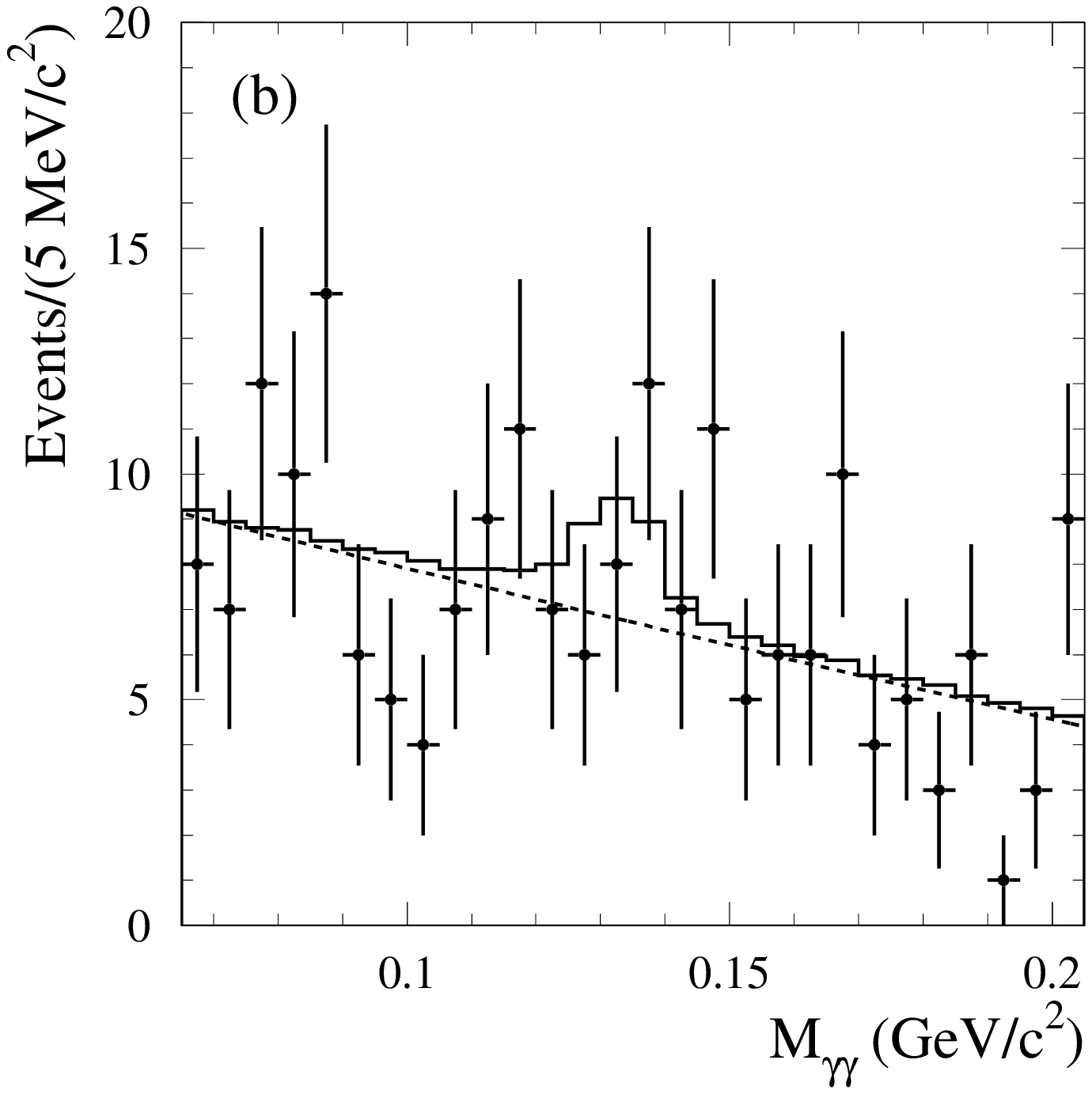}
\caption{
The two-photon invariant mass spectra for (a) $\eta$ and (b) 
$\eta^\prime$ events with two extra photons. The solid histograms 
represent the fit results. The dashed curves are the fitted distributions
for events without an extra $\pi^0$.
\label{fig8}}
\end{figure*}
Other possible sources of peaking background are the two-photon processes
$e^+e^- \to e^+e^- \eta^{(\prime)} \pi^0$. For the $\eta$ selection
the additional background comes from the two-photon production of 
$\eta^\prime$ mesons followed by the decay chain 
$\eta^\prime\to \pi^0\pi^0\eta$, $\eta\to \pi^+\pi^-\pi^0$. 
The $Q^2$ distribution of events from the latter background source is 
calculated from the $Q^2$ distribution of the selected $\eta^\prime$ events.
The ratio of the detection efficiencies for the two $\eta^\prime$ decay modes
is obtained from MC simulation. The total number of 
$\eta^\prime\to \pi^0\pi^0\eta$ events in the $\eta$ data sample is estimated
to be $17\pm2$. The events are concentrated almost entirely in the three 
lowest $Q^2$ bins.

To estimate background contributions from the 
$e^+e^- \to e^+e^- \eta^{(\prime)} \pi^0$ 
processes, we select events with two extra photons that each have an energy 
greater than 70 MeV. The distributions of the invariant mass of these extra 
photons for $\eta$ and $\eta^\prime$ events are shown in Fig.~\ref{fig8}. 
The invariant masses of the $\eta$ and $\eta^\prime$ candidates
are required to be in the mass windows 0.530--0.565 GeV/$c^2$ and
0.945--0.970 GeV/$c^2$, respectively. The spectra are fit by a sum
of the $\pi^0$ line shape obtained from simulated  
$e^+e^-\to e^+e^-\eta^{(\prime)}\pi^0$ events and a quadratic polynomial. 
The fitted numbers of events with an extra $\pi^0$ are $90\pm 20$ and 
$13\pm14$ for the $\eta$ and $\eta^\prime$ selections, respectively. 
It is expected that 8 events with an extra $\pi^0$ in the $\eta$ sample arise
from two-photon $\eta^\prime$ production. 

\begin{figure}
\includegraphics[width=.4\textwidth]{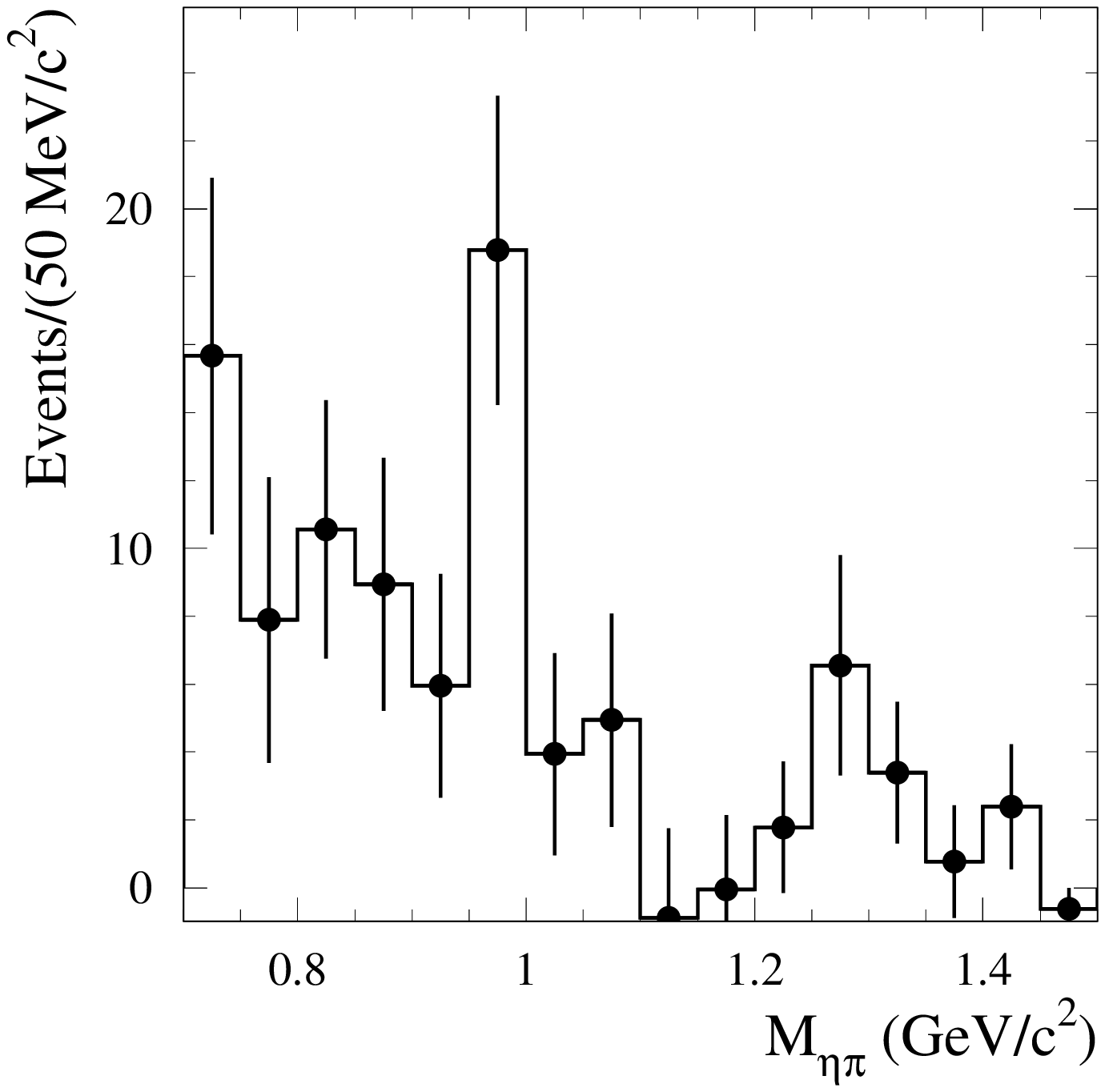}
\caption{The distribution of the $\eta\pi^0$ invariant mass for $\eta$ events 
with an extra $\pi^0$. The background from non-$\eta\pi^0$ events is subtracted.
\label{fig9}}
\end{figure}
The distribution of the 
$\eta\pi^0$ invariant mass for events with an extra $\pi^0$ is shown in 
Fig.~\ref{fig9}. The two-photon invariant mass of
the $\pi^0$ candidate is required to be in the 0.115--0.150 GeV/$c^2$ range.
The sidebands, 0.065--0.100 and 0.170--0.205 GeV/$c^2$, are used to subtract 
contamination from non-$\eta\pi^0$ events. 
It is known from two-photon measurements in the no-tag mode~\cite{2gtoetapi}
that the $\eta\pi^0$ final state is produced mainly via $a_0(980)$ and 
$a_2(1320)$
intermediate resonances. Evidence for these two intermediate resonances
is seen in the mass spectrum of Fig.~\ref{fig9}. 
Our spectrum differs significantly
from the spectrum for the no-tag mode~\cite{2gtoetapi}, which
is dominated by $a_2(1320)$ production.
In the no-tag mode the $a_2(1320)$ meson is produced predominantly in
a helicity-2 state, and thus with an angular distribution proportional to
$\sin^4{\theta_\pi}$,
where $\theta_\pi$ is the angle between the $\pi^0$ direction and the 
$\gamma\gamma$ collision axis in the $\gamma\gamma$ c.m.\ frame. Our 
selection criteria favor events with values of $\theta_\pi$ near zero 
and hence suppress helicity-2 states.

From MC simulation we estimate that the ratio of the number of 
$e^+e^-\to e^+e^-\eta^{(\prime)}\pi^0$ 
events with a detected $\pi^0$ to the number selected with 
standard criteria is about 2.5. 
For the $e^+e^-\to e^+e^-\eta^\prime$ process the estimated two-photon 
background does not exceed 1.6\% of the total number of selected 
$\eta^\prime$ events at 90\% confidence level. This background level is
treated as a measure of the systematic uncertainty due to possible
two-photon background for the $e^+e^-\to e^+e^-\eta^\prime$ process.

For the $e^+e^-\to e^+e^-\eta$ process the two-photon background is about
10\% of the total number of selected $\eta$ events. 
It should be noted that in the CLEO publication~\cite{CLEO} on measurements of
the meson-photon transition form factors the background from the two-photon 
production of the $\eta\pi^0$ final state was not considered.

A similar technique is used to estimate background
from the process $e^+e^-\to e^+e^-\phi,\,\phi\to \eta\gamma$
We do not see any $\phi$ meson signal in the
$\eta\gamma$ mass spectrum and estimate that this background   
does not exceed 10\% of the $\eta\pi^0$ background. The
$\eta\gamma$ events have the $r$ distribution similar to
that for $\eta\pi^0$ events, and are effectively subtracted
by the procedure described in the next section. The background
contributions from the processes
$e^+e^-\to e^+e^-\phi,\,\phi\to \eta^\prime\gamma$
is negligible due to the small $\phi\to \eta^\prime gamma$
branching fraction. The background from
$e^+e^-\to e^+e^- J/\psi,\,J/\psi \to \eta^{(\prime)}\gamma$
is estimated using the $Q^2$ distribution of
$e^+e^-\to e^+e^- J/\psi$ events measured in
Ref.~\cite{etacff} and efficiencies from MC simulations, and is
found to be negligible.

\subsection{Background subtraction from the $\eta$ data sample}\label{background3}
\begin{figure*}
\includegraphics[width=.4\textwidth]{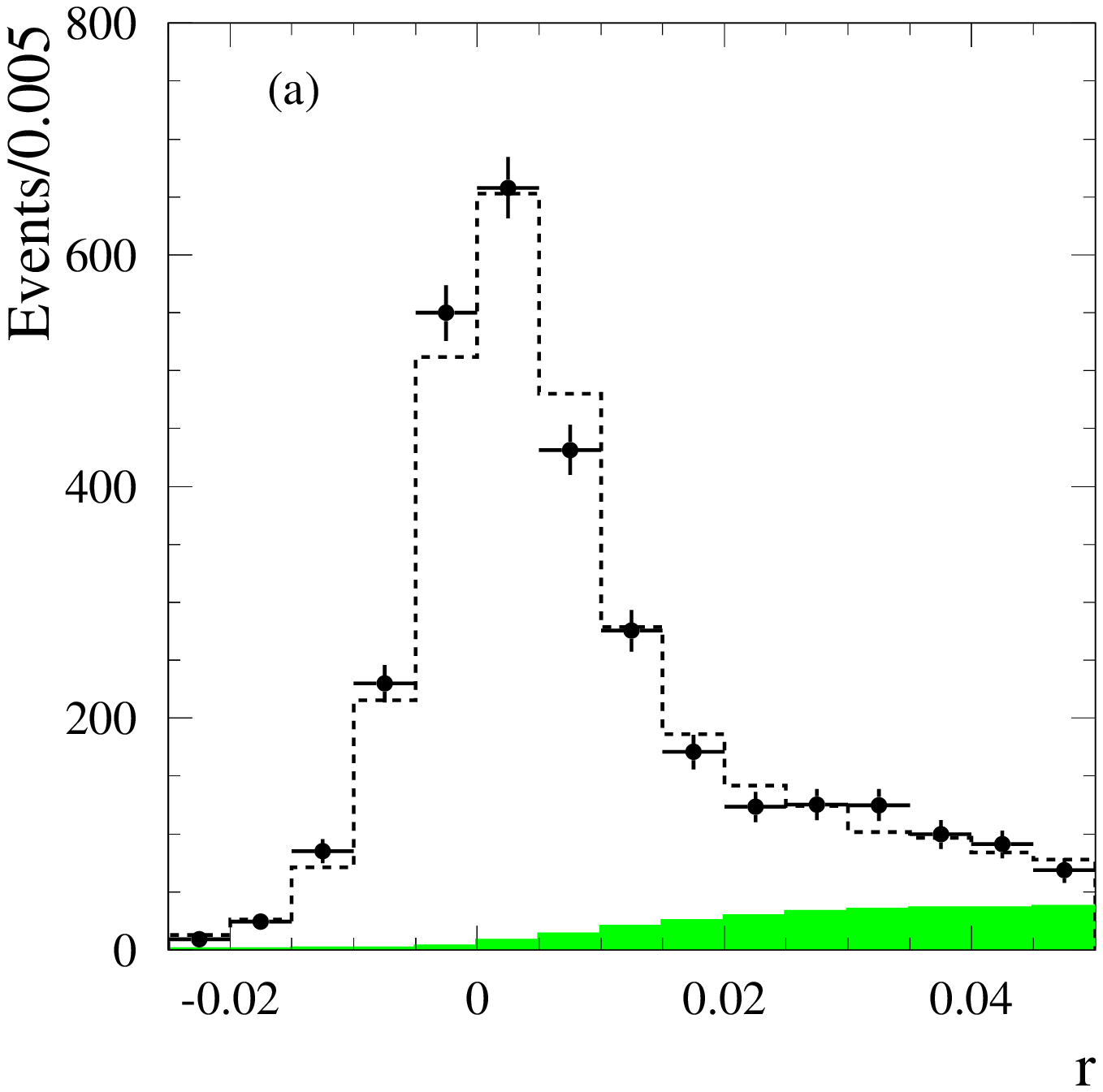}
\includegraphics[width=.4\textwidth]{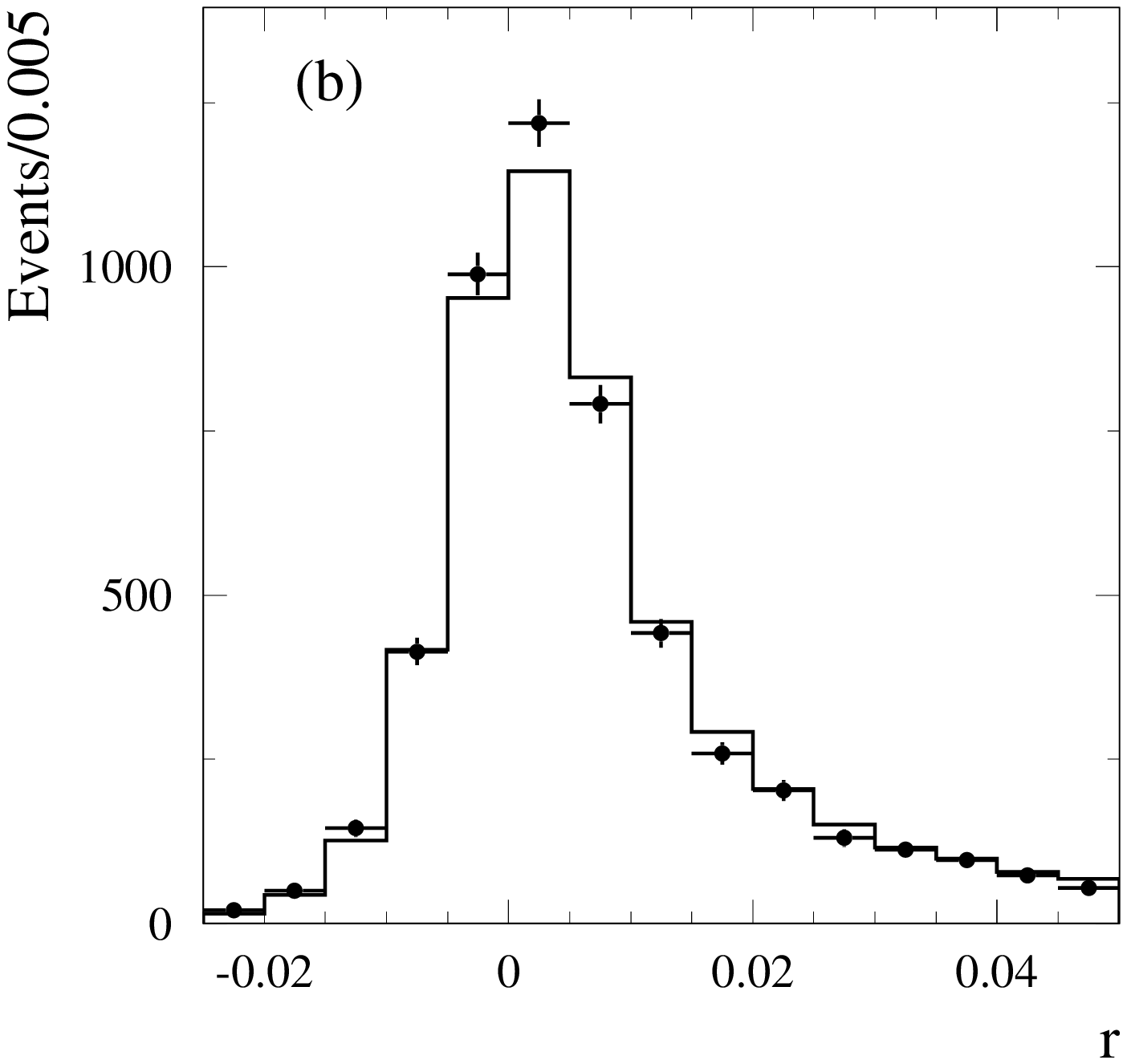}
\caption{
(a) The $r$ distribution for data events containing an $\eta$ (points with
error bars). The dashed histogram shows the fit results. The shaded histogram
is the fitted background contribution from the processes $e^+e^-\to e^+e^-\eta\pi^0$
and  $e^+e^-\to e^+e^-\eta^\prime\to e^+e^-\pi^0\pi^0\eta$. 
(b) The $r$ distribution for data events containing an $\eta^\prime$ 
(points with error bars). The solid histogram is the simulated distribution
for events from the signal  $e^+e^-\to e^+e^-\eta^\prime$ process normalized 
to the number of data events.
\label{fig10}}
\end{figure*}
To subtract background from the $\eta$ data sample the difference between 
the $r$ distributions for signal and background events is used.
The parameter $r$ is proportional
to the difference between the energy and the momentum of particles
recoiling against the $e\eta^{(\prime)}$ system and, therefore, is close to
zero for signal and has nonzero positive value for background events.
To obtain the $r$ distribution, data events are divided into 15 $r$ intervals. 
For each interval, the fit to the $\pi^+\pi^-\pi^0$ ($\pi^+\pi^-\eta$) spectra
is performed and the number of events containing an $\eta^{(\prime)}$ is 
determined. The $r$ distributions for events
in the $\eta$ and $\eta^\prime$ data samples are shown in Fig.~\ref{fig10}.

For $\eta^\prime$ events, for which the background is small, the data 
distribution is compared 
with the simulated signal distribution normalized to the number of data 
events. The distributions are in reasonable agreement. 
The ratio $R_s$ of the number of events with $r >0.025$
to the number with $r < 0.025$ is found to be $0.103\pm0.006$ in data and 
$0.116\pm0.002$ in simulation; the 13\% difference is taken as a
systematic uncertainty on the $R_s$ value for $\eta^\prime$ events
determined from simulation. Since the simulated $r$ distributions 
for $\eta^\prime$ and $\eta$ events are very close, the same systematic error
can be applied to $R_s$ value for $\eta$ events.

For $\eta$ events, the data $r$ distribution is fit with the sum of the
simulated distributions for signal and background $e^+e^-\to e^+e^-\eta\pi^0$
and $e^+e^-\to e^+e^-\eta^\prime\to e^+e^-\pi^0\pi^0\eta$ 
events. The fitted number of background events is $280\pm40$, in reasonable
agreement with the estimate given in the previous subsection based on the 
number of events with a detected extra $\pi^0$.

To subtract the background in each $Q^2$ interval the following
procedure is used. In Sec.~\ref{fitting} we described how the number of 
events containing an $\eta$ is determined for two regions of the parameter 
$r$: $-0.025 < r < 0.025$ ($N_1$) and $0.025 < r < 0.050$ ($N_2$). 
The numbers of signal and background events are then calculated as follows:
\begin{equation}
N_s=\frac{(1+R_s)(N_1 R_b-N_2)}{R_b-R_s}, \label{eqns}
\end{equation}
\begin{equation}
N_b=\frac{(1+R_b)(N_2-N_1 R_s)}{R_b-R_s}, \label{eqnb}
\end{equation}
where $R_s$ ($R_b$) is the $N_2/N_1$ ratio obtained from signal (background)
MC simulation. The expressions in Eqs.~(\ref{eqns}) and (\ref{eqnb}) are 
equivalent to a two-$r$-bin fit of data to signal and background MC 
predictions; fits using a higher 
number of bins are not useful due to lack of statistics.

The parameter $R_s$ is found to vary from 0.15 to 0.10 with 
increasing $Q^2$. The systematic uncertainty on $R_s$
(13\%) was estimated above.
To calculate $R_b$ for the $e^+e^-\to e^+e^-\eta\pi^0$ process, the simulated
background events are reweighted to reproduce the $\eta\pi^0$
mass spectrum observed in data (Fig.~\ref{fig9}). The $R_b$ value
varies from 2.0 to 1.5.  The systematic uncertainty on $R_b$ is estimated
based on its  $\eta\pi^0$ mass  dependence.  The maximum deviation
from the value averaged over the $\eta\pi^0$ spectrum of about 25\% is found 
when we exclude events with mass near the $\eta\pi^0$ threshold. 
This deviation is taken as an estimate of the systematic uncertainty on $R_b$.
The $r$ distribution for background events from two-photon $\eta^\prime$
production ($R_b$ is about 10) differs significantly from the distribution for 
$\eta\pi^0$ events. Therefore
we first subtract the calculated  $\eta^\prime$ contribution from $N_1$ and 
$N_2$ in each  $Q^2$ interval, and then calculate $N_s$ assuming that the
remaining background comes from the $e^+e^-\to e^+e^-\eta\pi^0$ process.
The obtained numbers of signal and 
background events are listed in Table~\ref{tab10}. The background
includes both the $e^+e^-\to e^+e^-\eta\pi^0$ and $e^+e^-\to e^+e^-\eta^\prime$
contributions.
The systematic errors quoted for $N_s$ are mainly due to the uncertainties on 
$R_s$ and $R_b$. 

\section{Detection efficiency\label{deteff}}
\begin{figure*}
\includegraphics[width=.4\textwidth]{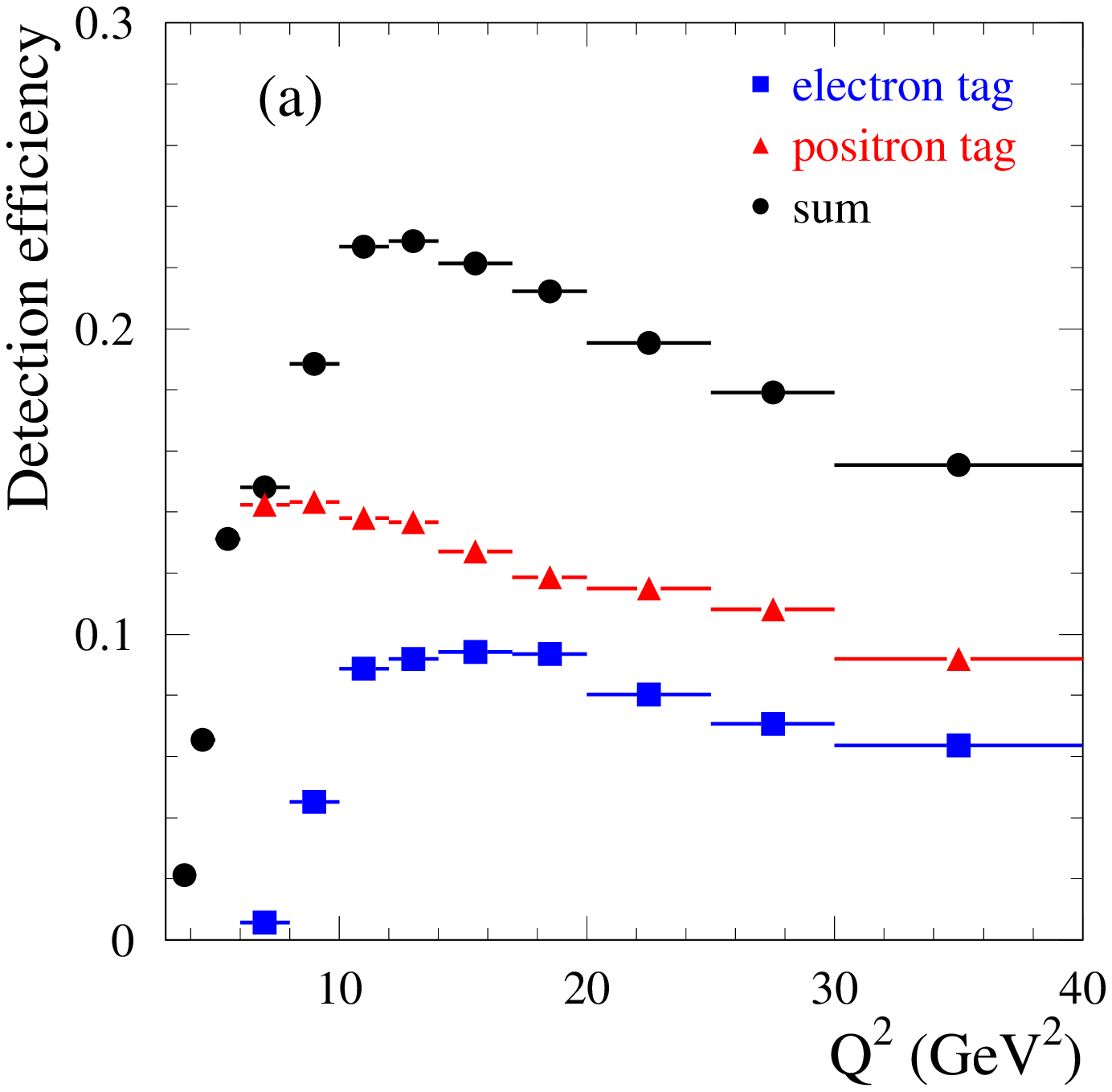}
\includegraphics[width=.4\textwidth]{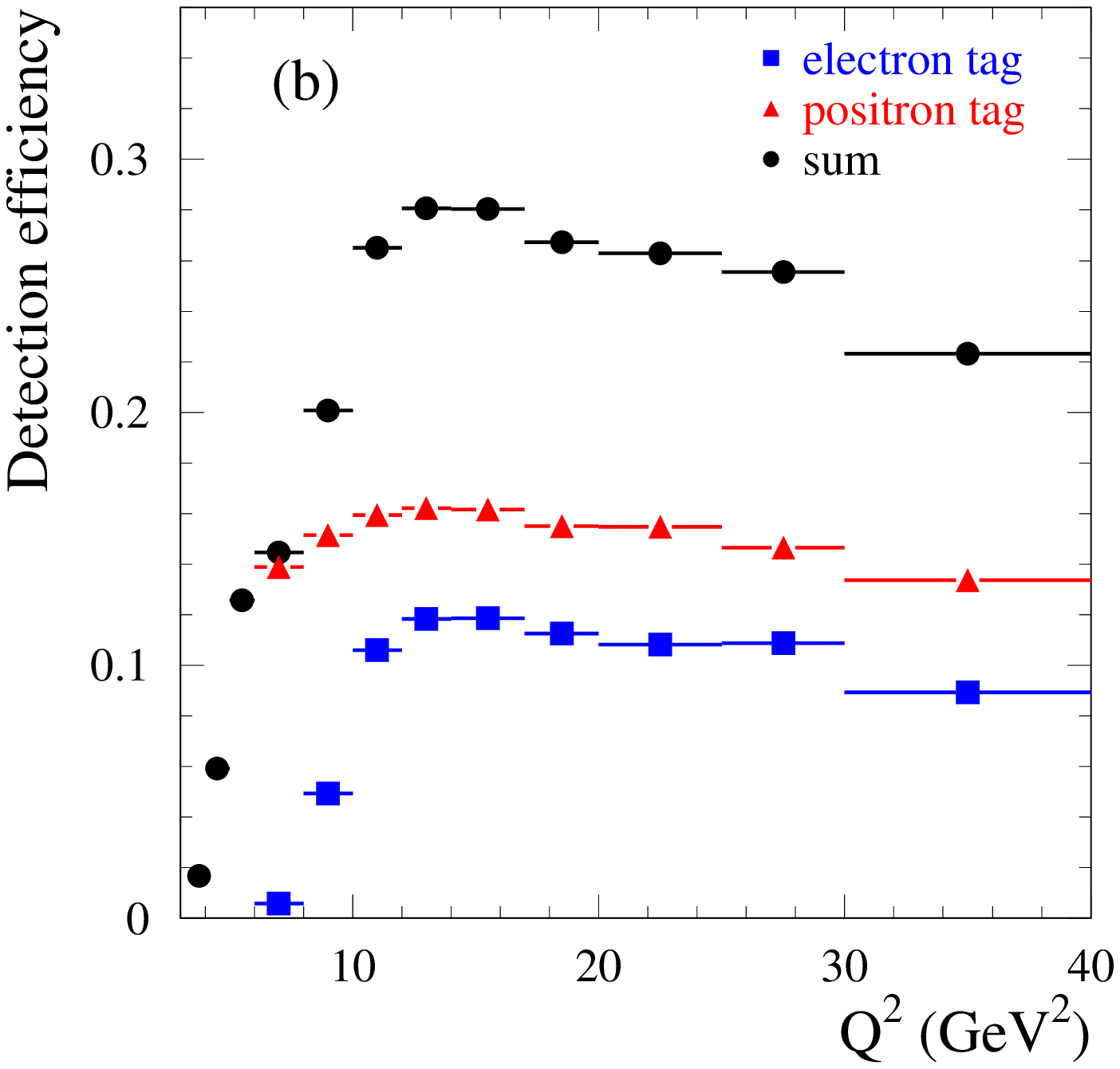}
\caption{The detection efficiencies for (a) $e^+e^-\to e^+e^-\eta$ 
with $\eta\to \pi^+\pi^-\pi^0$ and $\pi^0\to 2\gamma$
and (b) $e^+e^-\to e^+e^-\eta^\prime$ with $\eta\to \pi^+\pi^-\eta$
and $\eta\to 2\gamma$ as functions of 
the momentum transfer squared for events with a tagged electron 
(squares), a tagged positron (triangles), and their sum (circles).
In the region  $Q^2 < 6$ GeV$^2$, where the electron-tag efficiency is close to zero, 
the sum and the positron-tag efficiencies coincide.
\label{fig11}}
\end{figure*}
The detection efficiency is determined from MC simulation as the ratio
of the true $Q^2$ distributions computed after and before applying the
selection criteria. The $Q^2$ dependencies of the detection efficiencies
for both processes under study are shown in Fig.~\ref{fig11}. 
The detector acceptance limits the detection
efficiency at small $Q^2$. 
The cross sections are measured in the regions $Q^2 > 4$~GeV$^2$, where the 
detection efficiencies are greater than 5\%. The asymmetry of the $e^+e^-$ collisions
at PEP-II leads to different efficiencies for events with electron and 
positron tags. The $Q^2$ range from 4 to 6 GeV$^2$ is measured only with
the positron tag. 

We study possible sources of systematic uncertainty due to differences
between data and MC simulation in detector response. 
The MC simulation predicts about a 2.5\% loss of signal events, weakly 
dependent on $Q^2$, 
due to the offline trigger, i.e. program filters, 
which provide background suppression before the full event reconstruction.
Events of the process under study satisfying our selection criteria
pass a filter selecting events
with at least three tracks in the drift chamber originating from the
interaction region.
The filter inefficiency is measured from data using 
a small fraction of selected events that 
does not pass the background filters. 
Combining events from the $\eta$ and $\eta^\prime$ samples, we determine
the ratio of the inefficiencies in data and MC
simulation to be $1.15\pm0.20$. The error of the ratio
is used to estimate the systematic uncertainty 
for the filter inefficiency: $0.2\times 2.5=0.5\%$.
The trigger inefficiency obtained using MC simulation is 
about 1\% in the first $Q^2$ interval (4--5 GeV$^2$) and falls to 
zero at  $Q^2 > 14$ GeV$^2$. 
The limited statistics do not allow us to measure this inefficiency
in data. Therefore, the level of the inefficiency observed in
the MC simulation is taken as an estimate of the systematic 
uncertainty due to the trigger inefficiency.

The systematic uncertainty due to a possible difference between data
and simulation in the charged-particle track reconstruction for 
pions is estimated to be about 0.35\% per track, 
so the total uncertainty is 0.7\%. 
For electron tracks, this uncertainty is about 0.1\%.

The data-MC simulation 
difference in the pion identification efficiency is estimated using 
the identification 
efficiencies measured for pions in the 
$D^{\ast +}\to D^0\pi^+,\;D^0\to\pi^+ K^-$ decay. The ratio of
the data and MC identification efficiencies is determined as a function
of the pion momentum and polar angle. These functions for positive and
negative pions are then convolved with the pion energy and angular distributions
for simulated signal events in each $Q^2$ interval. 
The resulting efficiency correction ($\delta_{\pi}$) for pion identification varies 
from -1\% to 0.5\% in the $Q^2$ range from 4 to 40 GeV$^2$. The systematic uncertainty
in the correction does not exceed 0.5\%.
The data-MC simulation difference in electron identification is estimated
using the identification efficiencies measured for electrons in 
radiative Bhabha events. The found efficiency correction ($\delta_{e}$) does not 
exceed 1\%. Its systematic uncertainty is estimated to be 0.5\%.

\begin{figure}
\includegraphics[width=.4\textwidth]{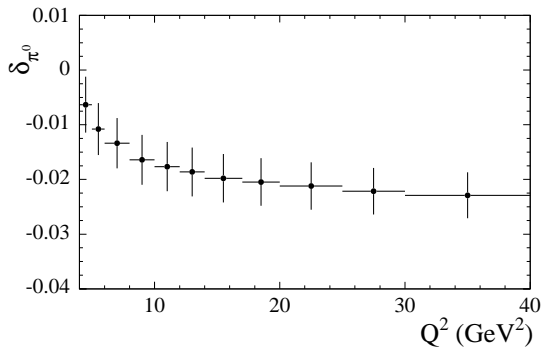}
\caption{The correction to the MC-estimated $\pi^0$ reconstruction
efficiency $\delta_{\pi^0}$ as a function of $Q^2$ for the $e^+e^-\to e^+e^-\eta$ 
process.
\label{fig12}}
\end{figure}
The $\pi^0$ reconstruction efficiency is studied 
using events from the ISR process $e^+e^-\to \gamma\omega$, 
$\omega\to\pi^+\pi^-\pi^0$. 
These events can be reconstructed and selected
without using information related to the $\pi^0$.
The $\pi^0$ reconstruction efficiency is computed as the
ratio of the number of events with an identified $\pi^0$ to
the total number of reconstructed $e^+e^-\to \gamma\omega$ events.
The data-MC simulation relative difference
in the $\pi^0$ efficiency 
depends on the $\pi^0$ momentum and 
varies from $(0.7\pm 1.2)\%$ at momenta below 0.25 GeV/$c$ to 
$(-4.2\pm1.3)\%$ at 4~GeV/$c$~\cite{pi0ff}. The efficiency correction averaged 
over the $\pi^0$ spectrum is shown in Fig.~\ref{fig12} as a
function of $Q^2$. The systematic uncertainty associated with this
correction is estimated to be 1\%. For $\eta\to \gamma\gamma$
decays the efficiency correction is expected to be smaller.
The maximum value of the $\pi^0$  efficiency correction (2\%) is
conservatively taken as an estimate of systematic uncertainty
due to a possible data-MC simulation difference in 
the $\eta\to \gamma\gamma$ reconstruction.

To estimate the effect of the requirement $-0.025<r<0.05$,  
$\eta^\prime$ events with $0.05<r<0.075$ are studied. 
We calculate the double ratio minus unity
\begin{equation}
\frac{\Delta\sigma}{\sigma}=
\frac{(N_{\rm new}/N)_{\rm data}}{(N_{\rm new}/N)_{\rm MC}}-1,
\label{drat}
\end{equation}
where  $N_{new}$ and $N$ are the numbers of signal events with the new
and standard selection criteria. The ratio is sensitive to the relative change
in the measured cross section due to the changes in the selection criteria.
We do not observe any significant $Q^2$ dependence of 
$\Delta{\sigma}/{\sigma}$. The average over $Q^2$ is found to be 
consistent with zero ($-0.003\pm0.004$). We conclude that the simulation 
reproduces the shape of the $r$ distribution.

We also study the effect of the $|\cos{\theta^\ast_{e\eta^{(\prime)}}}| > 0.99$ 
restriction by changing the value to 0.95. The corresponding change
of the measured cross section does not depend on $Q^2$.
The average change in cross section integrating 
over $Q^2$ is $(2.0\pm0.4)\%$.
We consider this data-MC simulation difference (2\%) as a measure
of the systematic uncertainty due to the $\cos{\theta^\ast_{e\eta(^\prime)}}$
criterion.

\begin{figure}
\includegraphics[width=.4\textwidth]{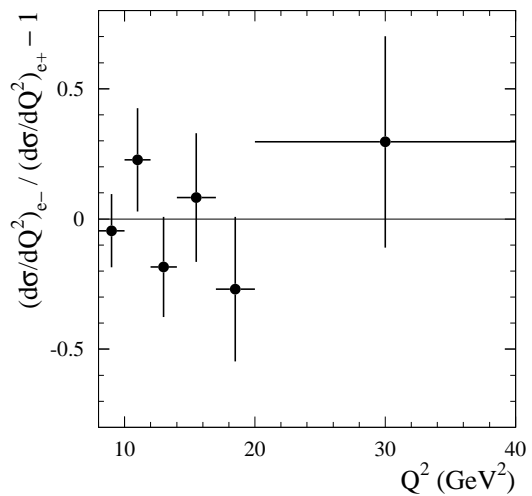}
\caption{The ratio of the cross sections for the $e^+e^-\to e^+e^-\eta$ process
measured with electron and positron tags as a function of $Q^2$.
\label{fig13}}
\end{figure}
The angular and energy distributions of detected particles are very different for
events with electron and positron tags. As a cross-check of our study of the efficiency
corrections, we have performed comparison of $Q^2$ dependencies of the cross
sections obtained with only electron and only positron tags. For $Q^2>8$ GeV$^2$,
where both positron and electron data are available, the ratio of the cross sections
have been found to be consistent with unity, for both $\eta$ and $\eta^\prime$ events.
The $Q^2$ dependence of the ratio for $\eta$ events is shown in Fig.~\ref{fig13}. 
Due to limited statistics data of the three highest $Q^2$ bins are combined. 

The main 
sources of systematic uncertainty associated with the detection efficiency 
are summarized in Table~\ref{tab20} for both processes under study.
The values of the detection efficiency and the total efficiency correction
$\delta_{{\rm total}}=\delta_{\pi}+\delta_{e}+\delta_{\pi^0}$ 
(the term $\delta_{\pi^0}$ is only applicable to the $\eta$ mode)
for different $Q^2$ intervals are listed in Tables~\ref{tab10} and
~\ref{tab11}. The data distribution is corrected as follows:
\begin{equation}
N^{corr}_i=N_i/(1+\delta_{{\rm total},i}),
\label{eqcor}
\end{equation}
where $N_i$ is the number of signal events in the $i$th $Q^2$ interval. 
\begin{table}
\caption{The main sources of systematic uncertainty associated with the
detection efficiency, and the total efficiency systematic uncertainty for 
$e^+e^-\to e^+e^-\eta$ and $e^+e^-\to e^+e^-\eta^\prime$ events.
\label{tab20}}
\begin{ruledtabular}
\begin{tabular}{lcc}
Source & $\eta$ (\%) & $\eta^\prime$ (\%) \\
\hline
Track reconstruction                                      & \multicolumn{2}{c}{0.8} \\
$\pi^\pm$ identification                                  & \multicolumn{2}{c}{0.5} \\
$e^\pm$ identification                                    & \multicolumn{2}{c}{0.5} \\
$|\cos{\theta^\ast_{e\eta^{(\prime)}}}| > 0.99$ criterion & \multicolumn{2}{c}{2.0} \\
Trigger, filters                                          & 0.7        & 1.3       \\
$\eta,\pi^0\to2\gamma$ reconstruction                     & 1.0        & 2.0            \\
\hline
Total                                                     & 2.6        & 3.3       \\
\end{tabular}
\end{ruledtabular}
\end{table}

\section{Cross section and form factor}\label{crosssec}
The Born differential cross section for 
$e^+e^-\to e^+e^-\eta^{(\prime)}$ is 
\begin{equation}
\frac{{\rm d} \sigma}{{\rm d} Q^2}=
\frac{({\rm d}N/{\rm d}Q^2)_{\rm corr}^{\rm unfolded}}{\varepsilon RLB}
\label{eqcs}
\end{equation}
where $({\rm d}N/{\rm d}Q^2)_{\rm corr}^{\rm unfolded}$ is the mass spectrum
corrected for
data-MC simulation differences and unfolded for detector resolution effects as 
explained below, $L$ is the total integrated luminosity, $\varepsilon$ is the 
$Q^2$-dependent detection efficiency, and $R$ is a radiative correction factor
accounting for distortion of the $Q^2$ spectrum due to vacuum polarization
effects and the emission of soft photons from the initial-state particles.
The factor $B$ is the product of the branching fractions, 
${\cal B}(\eta \to \pi^+\pi^-\pi^0){\cal B}(\pi^0 \to \gamma\gamma)=0.2246\pm0.0028$ or
${\cal B}(\eta^\prime \to \pi^+\pi^-\eta){\cal B}(\eta \to \gamma\gamma)=0.1753\pm0.0056$~\cite{pdg}.

The radiative correction factor  $R$ is determined using simulation
at the generator level, i.e., without detector simulation. The $Q^2$ spectrum
is generated using only the pure Born amplitude for the 
$e^+e^-\to e^+e^-\eta^{(\prime)}$ process, and then using a model with radiative 
corrections included. The radiative correction factor, 
evaluated as the ratio of the second spectrum to the first, varies from
0.994 at $Q^2=4$ GeV$^2$ to 1.002 at $Q^2=40$ GeV$^2$.
The accuracy of the radiative correction calculation is estimated to be 
1\%~\cite{RC}.
It should be noted that the value of $R$ depends on the requirement on
the extra photon energy. The $Q^2$ dependence obtained corresponds to
the condition $r=2E^\ast_\gamma/\sqrt{s}<0.1$ imposed in the simulation.

The corrected and unfolded $Q^2$ distribution
$({\rm d}N/{\rm d}Q^2)_{\rm corr}^{\rm unfolded}$
is obtained from the measured distribution 
by dividing by the efficiency correction factor (see Eq.(\ref{eqcor})) and
unfolding for the effect of finite $Q^2$ resolution. Using MC simulation,
a migration matrix $H$ is obtained, which represents the probability that
an event with true $Q^2$ in interval $j$ is reconstructed in interval $i$:
\begin{equation}
\left( \frac{{\rm d}N}{{\rm d}Q^2} \right)^{\rm rec}_i= 
\sum_{j}H_{ij}\left( \frac{{\rm d}N}{{\rm d}Q^2} \right)^{\rm true}_j.
\end{equation}                                                                                    
In the case of extra photon emission, $Q^2_{\rm true}$ is calculated
as $-(p-p^\prime-k)^2$, where $k$ is the photon four-momentum;
$\varepsilon$ and $R$ in Eq.(\ref{eqcs}) are functions of $Q^2_{\rm true}$.
As the chosen $Q^2$ interval width significantly exceeds the resolution
for all $Q^2$, non-zero elements of the  migration matrix lie 
on and near the diagonal. The values of the diagonal elements are in the
range 0.9--0.95.                        
The true $Q^2$ distribution is obtained by applying the inverse of 
the migration matrix to the measured distribution. The procedure does 
not change the shape of the $Q^2$ distribution significantly, but increases
the errors (by about 10\%) and their correlations. The number of events
($N_{\rm corr}^{\rm unfolded}$) as a function of  $Q^2$ is reported in
Tables~\ref{tab10} and \ref{tab11}. 
  
The value of the differential cross section as a function of  $Q^2$ is listed
in Tables~\ref{tab30} and \ref{tab31}. The quoted errors are statistical and 
systematic. The latter includes only $Q^2$-dependent errors: the systematic 
uncertainty in the number of signal events and the statistical errors on the 
efficiency correction and MC simulation. The $Q^2$-independent systematic error
on the $e^+e^-\to e^+e^-\eta$ cross section
is 3.5\%; this results from the uncertainties on the detection 
efficiency, both systematic (2.6\%) and model-dependent (1.5\%),
the uncertainty in the calculation of the radiative correction factor (1\%), 
and the errors on the integrated luminosity (1\%) and the $\eta$ decay 
branching fraction (1.2\%)~\cite{pdg}.
The $Q^2$-independent systematic error on the $e^+e^-\to e^+e^-\eta^\prime$ 
cross section is 5.3\%. It includes the systematic and model uncertainties 
on the detection efficiency (3.3\% and 1.5\%, respectively), the uncertainties on 
the background subtraction (1.6\%) and the radiative correction factor (1\%), 
and the errors on the integrated luminosity (1\%) and the $\eta^\prime$ decay 
branching fraction (3.2\%)~\cite{pdg}.

The model dependence
of the detection efficiency arises from the unknown cross-section 
dependence on the momentum transfer to the untagged electron. 
The MC simulation is performed, and the detection efficiency
is determined, with the restriction that the momentum transfer
to the untagged electron be greater than $-0.6$ GeV$^2$, so that
the cross section is measured for the restricted range
$|q_2^2| < 0.6$ GeV$^2$. The actual $q_2^2$ threshold is determined
by the requirement on $\cos{\theta^\ast_{e\eta^{(\prime)}}}$ 
and is equal to 0.38 GeV$^2$. The MC simulation is performed 
with a $q^2_2$ independent form factor, which corresponds to the 
QCD-inspired model 
$F(q_1^2,q_2^2)\propto 1/(q_1^2+q_2^2) \approx 1/q_1^2$~\cite{n23}.
The event loss due to the $|q_2^2| < 0.38$ GeV$^2$ restriction is about 2.5\%.
The use of the form factor predicted by the vector dominance model 
$F(q_2^2)\propto 1/(1-q_2^2/m^2_\rho)$, where $m_\rho$ is $\rho$
meson mass, leads to a decreased event loss of only 1\%.
The difference between these efficiencies is considered to be
an estimate of the model uncertainty due to the unknown $q_2^2$ 
dependence.

\begin{figure*}
\includegraphics[width=.45\textwidth]{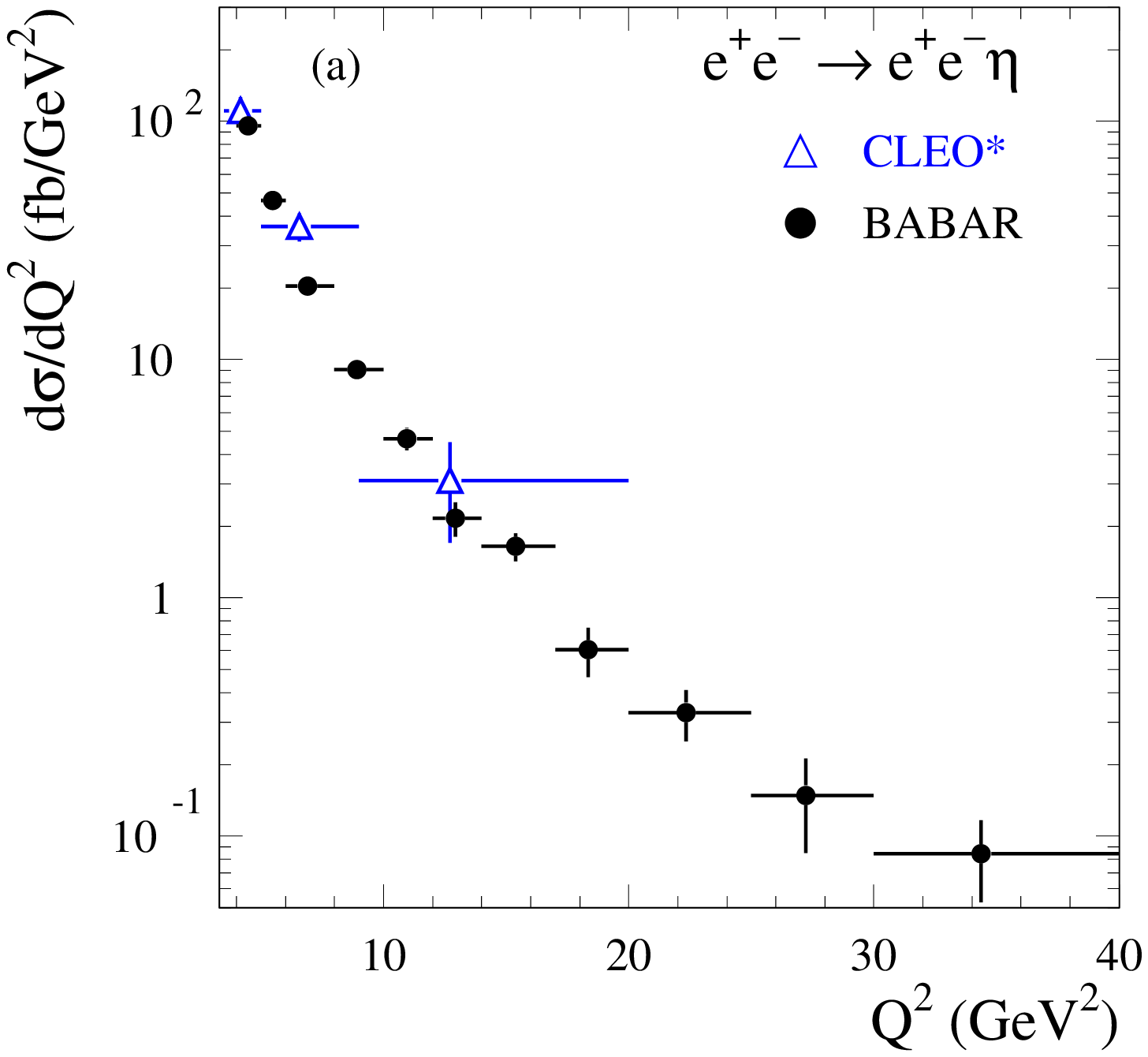}
\includegraphics[width=.45\textwidth]{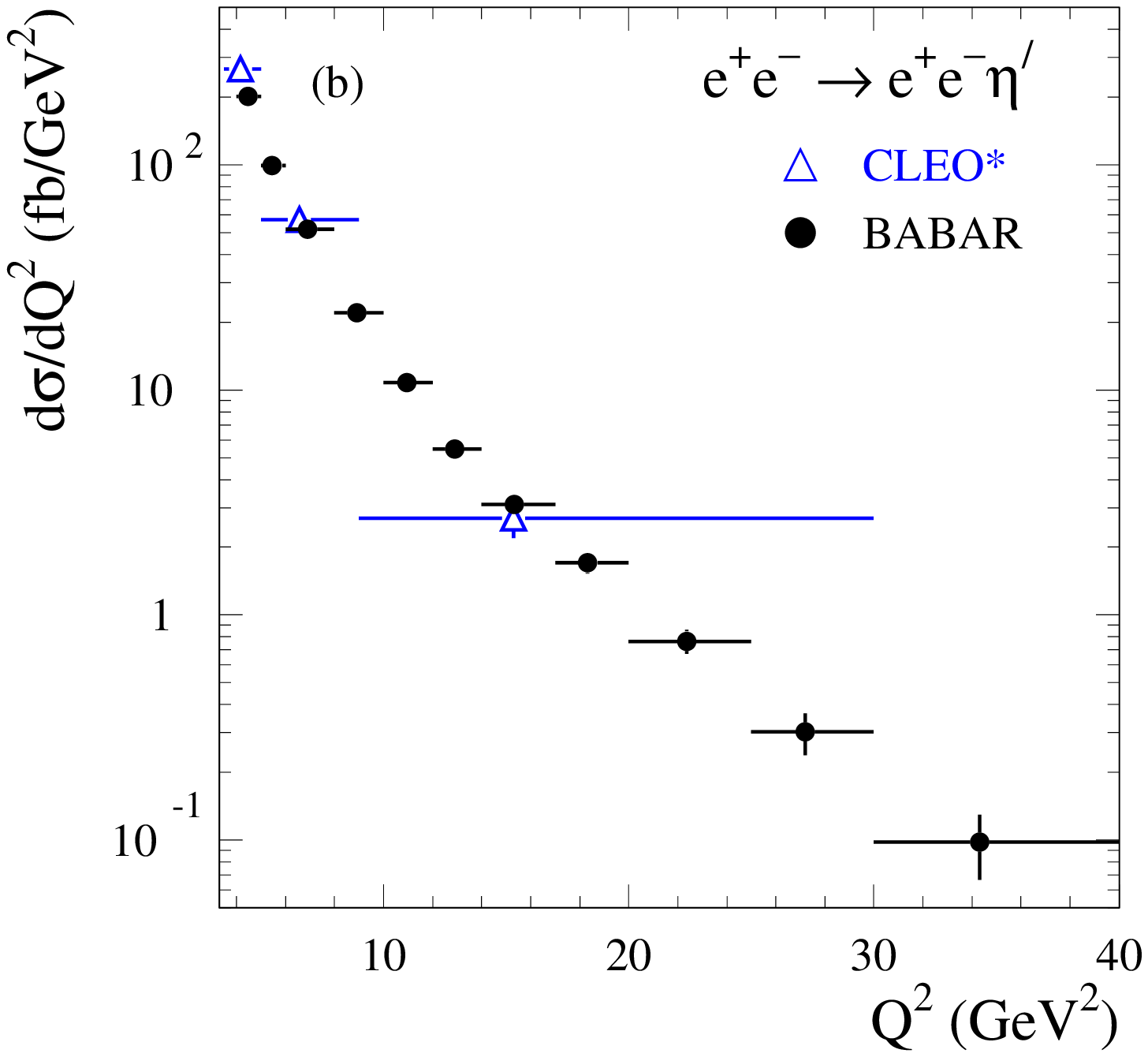}
\caption{The differential cross sections for 
(a) $e^+e^-\to e^+e^-\eta$ and (b) $e^+e^-\to e^+e^-\eta^\prime$
from the present analysis compared to those from the CLEO 
experiment~\cite{CLEO}. The asterisk near the label ``CLEO'' in this and
next figures indicates that the original CLEO results obtained
in different $\eta^{(\prime)}$ decay modes were averaged assuming that 
systematic errors for different modes are not correlated.
In the present analysis the cross sections are measured with the restriction
$|q_2^2| < 0.6$ GeV$^2$. In the CLEO analysis the cross sections have
been obtained using the vector dominance model for the $q_2^2$ dependence
in simulation.
\label{fig14}}
\end{figure*}

Because of the strong nonlinear dependence of the cross section 
on $Q^2$, the effective value of $Q^2$ corresponding to the measured 
cross section differs from the center of the
$Q^2$ interval. We parametrize the measured cross section with a smooth
function and calculate $\overline{Q^2}$ 
for each $Q^2$ interval solving the equation
$${\rm d}\sigma/{\rm d}(Q^2)(\overline{Q^2})=
{\rm d}\sigma/{\rm d}(Q^2)_{\rm average},$$
where ${\rm d}\sigma/{\rm d}(Q^2)_{\rm average}$ is the 
differential cross section averaged over the interval.
The values of $\overline{Q^2}$
are listed in Table~\ref{tab30} and Table~\ref{tab31}.
The measured differential cross sections for both processes under
study are shown in Fig.~\ref{fig14}, together with the data 
reported by the CLEO Collaboration~\cite{CLEO}
for $Q^2 > 3.5$ GeV$^2$. We average the CLEO results obtained
in different $\eta^{(\prime)}$ decay modes assuming that systematic errors for
different modes are not correlated.
\begin{table}
\caption{The $Q^2$ interval, the weighted average $Q^2$ value for the 
interval ($\overline{Q^2}$), the $e^+e^-\to e^+e^-\eta$ cross section
(${{\rm d}\sigma}/{{\rm d}Q^2}(\overline{Q^2})$), and the product of the              
$\gamma\gamma^\ast\to \eta$ transition form factor $F(\overline{Q^2})$
and $\overline{Q^2}$.  The statistical and systematic errors are quoted 
separately for the cross sections, and are combined for the form factors.
In the table we quote the $Q^2$-dependent systematic errors. The
$Q^2$-independent error is 3.5\% for the cross section and
2.9\% for the form factor.
\label{tab30}}
\begin{ruledtabular}
\begin{tabular}{cccccc}
$Q^2$ interval & $\overline{Q^2}$ &
${{\rm d}\sigma}/{{\rm d}Q^2}(\overline{Q^2})$ &
$\overline{Q^2}|F(\overline{Q^2})|$ \\
(GeV$^2$)      &     (GeV$^2$)    &                (fb/GeV$^2$)     & (MeV) \\
\hline 
  4--5  &  4.47 & $95.6\pm5.1\pm3.1     $ & $143.4\pm 4.4$ \\                                 
  5--6  &  5.47 & $46.6\pm2.7\pm1.7     $ & $142.7\pm 4.9$ \\                                 
  6--8  &  6.89 & $20.4\pm1.2\pm0.8     $ & $142.6\pm 5.2$ \\                                 
  8--10 &  8.92 & $9.06\pm0.72\pm0.35   $ & $151.2\pm 6.7$ \\                                 
 10--12 & 10.96 & $4.67\pm0.47\pm0.18   $ & $158.5\pm 8.5$ \\                                 
 12--14 & 12.92 & $2.16\pm0.34\pm0.10   $ & $146.5\pm12.1$ \\                                 
 14--17 & 15.38 & $1.65\pm0.22\pm0.06   $ & $178.9\pm12.1$ \\                                 
 17--20 & 18.34 & $0.61\pm0.14\pm0.02   $ & $151.6\pm17.8$ \\                                 
 20--25 & 22.33 & $0.33\pm0.08\pm0.01   $ & $166.0\pm20.2$ \\                                 
 25--30 & 27.23 & $0.15\pm0.06\pm0.01   $ & $166.7\pm36.6$ \\                                 
 30--40 & 34.38 & $0.085\pm0.032\pm0.003$ & $205.9\pm39.0$ \\
\end{tabular}
\end{ruledtabular}
\end{table}
\begin{table}
\caption{The $Q^2$ interval, the weighted average $Q^2$ value for the 
interval ($\overline{Q^2}$), the $e^+e^-\to e^+e^-\eta^\prime$ cross section
(${{\rm d}\sigma}/{{\rm d}Q^2}(\overline{Q^2})$), and the product of the              
$\gamma\gamma^\ast\to \eta^\prime$ transition form factor $F(\overline{Q^2})$
and $\overline{Q^2}$. The statistical and systematic errors are quoted 
separately for the cross sections, and are combined for the form factors.
In the table we quote the $Q^2$-dependent systematic errors. The
$Q^2$-independent error is 5.3\% for the cross section and
3.5\% for the form factor.
\label{tab31}}
\begin{ruledtabular}
\begin{tabular}{cccccc}
$Q^2$ interval & $\overline{Q^2}$ &
${{\rm d}\sigma}/{{\rm d}Q^2}(\overline{Q^2})$ &
$\overline{Q^2}|F(\overline{Q^2})|$ \\
(GeV$^2$)      &     (GeV$^2$)    &                (fb/GeV$^2$)     & (MeV) \\
\hline 
  4--5  &  4.48 & $202\pm 7\pm 3$           & $216.2\pm 4.3$ \\ 
  5--6  &  5.46 & $ 99.6\pm 3.6\pm 1.4$     & $214.3\pm 4.1$ \\ 
  6--8  &  6.90 & $ 51.7\pm 1.6\pm 0.5$     & $233.3\pm 3.9$ \\ 
  8--10 &  8.92 & $ 22.1\pm 0.9\pm 0.2$     & $241.6\pm 5.2$ \\ 
 10--12 & 10.95 & $ 10.8\pm 0.6\pm 0.1$     & $245.5\pm 6.7$ \\ 
 12--14 & 12.90 & $ 5.45\pm 0.41\pm 0.06$   & $236.7\pm 8.9$ \\ 
 14--17 & 15.33 & $ 3.10\pm 0.24\pm 0.04$   & $248.5\pm 9.9$ \\ 
 17--20 & 18.33 & $ 1.70\pm 0.18\pm 0.02$   & $258.7\pm13.7$ \\ 
 20--25 & 22.36 & $ 0.77\pm 0.09\pm 0.01$   & $257.0\pm15.4$ \\ 
 25--30 & 27.20 & $ 0.30\pm 0.06\pm 0.01$   & $240.0\pm25.7$ \\ 
 30--40 & 34.32 & $ 0.098\pm 0.031\pm 0.002$& $224.1\pm35.9$ \\
\end{tabular}
\end{ruledtabular}
\end{table}

\begin{figure*}
\includegraphics[width=.45\textwidth]{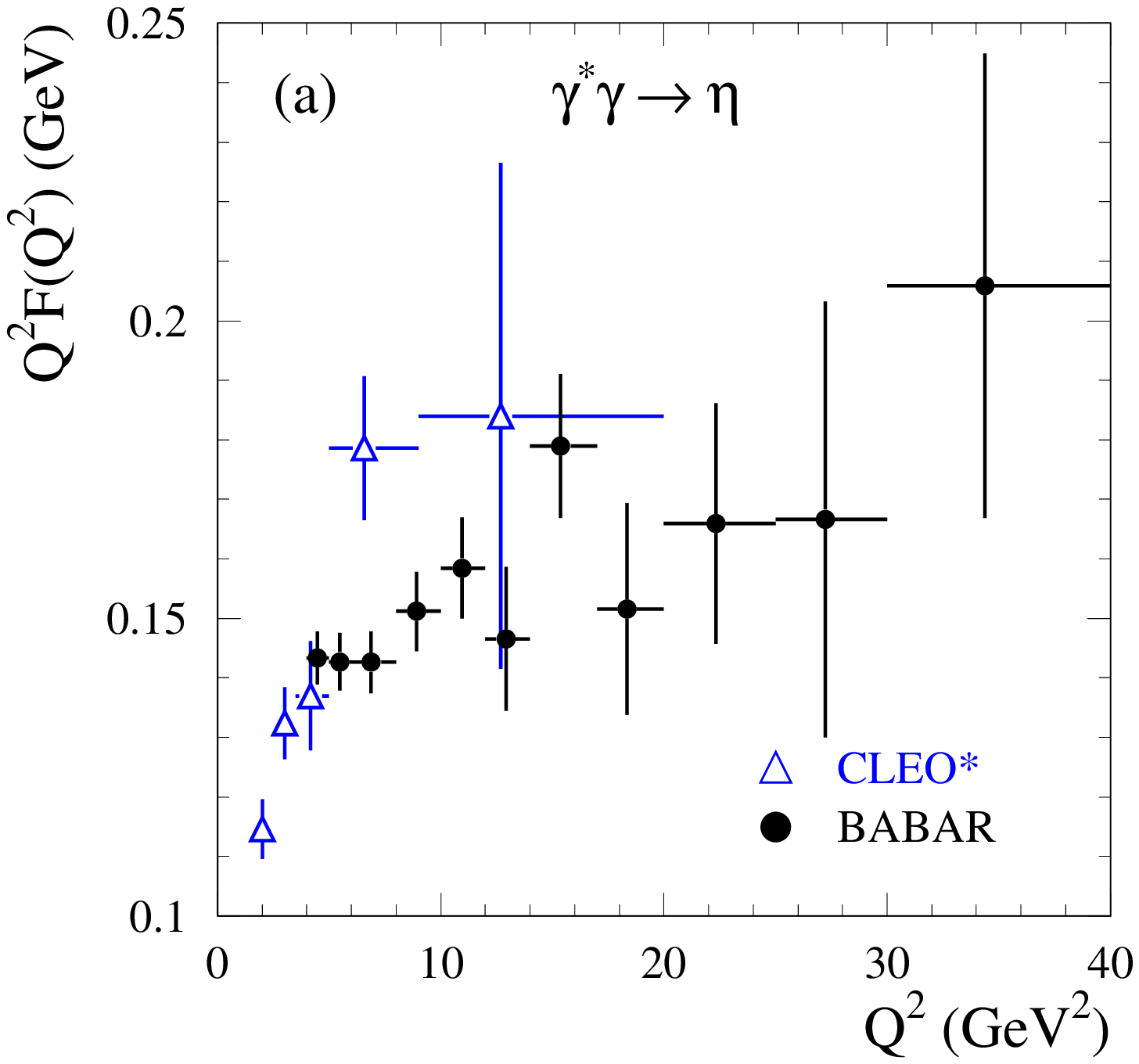}
\includegraphics[width=.45\textwidth]{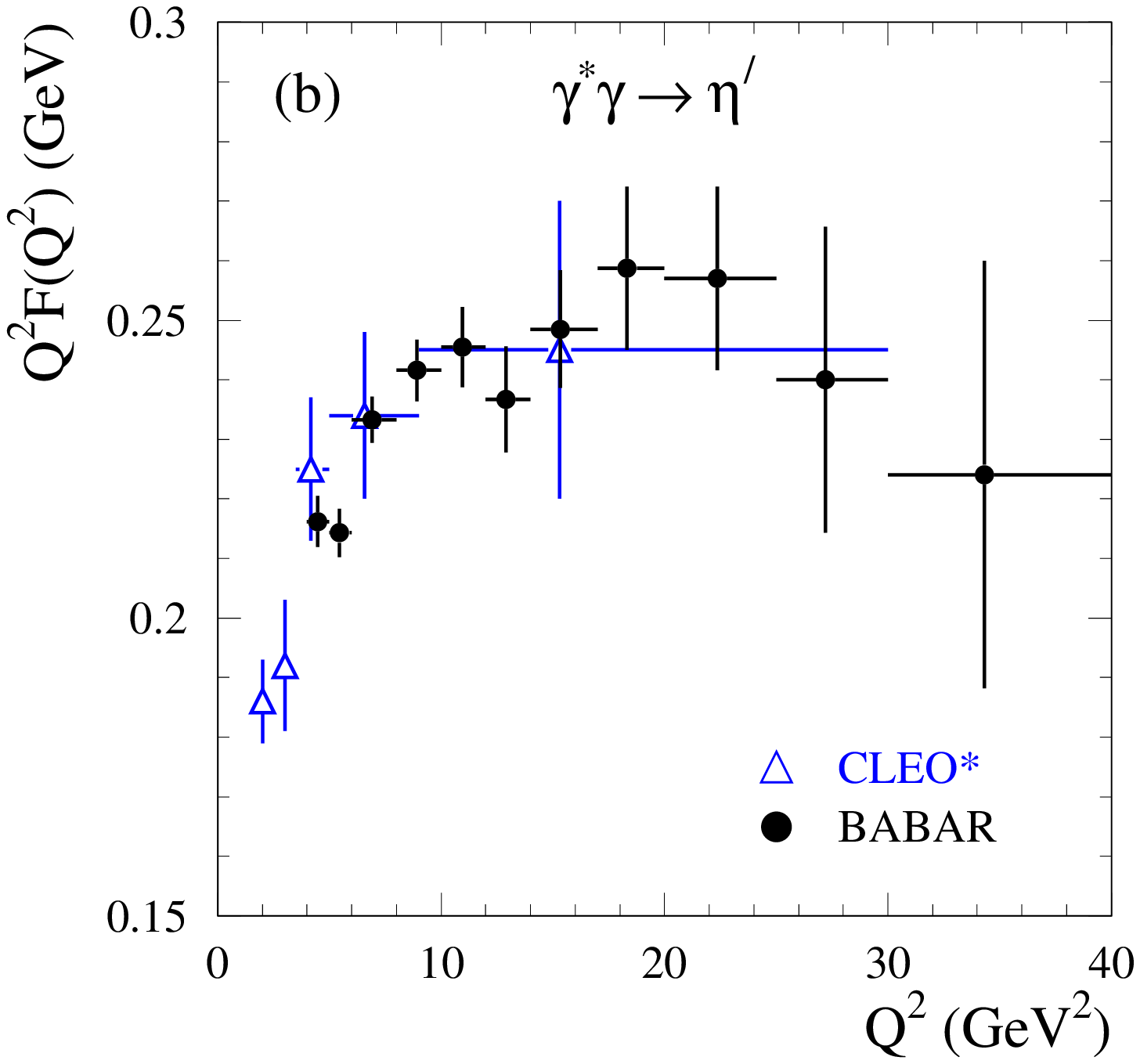}
\caption{The transition form factors multiplied by $Q^2$ for 
(a) $\gamma\gamma^\ast\to\eta$  and 
(b) $\gamma\gamma^\ast\to\eta^\prime$.
\label{fig15}}
\end{figure*}

To extract the transition form factor, the measured and calculated
cross sections are compared. The simulation uses a constant form 
factor $F^2_{\rm MC }$. Therefore, the measured form factor is 
determined from
\begin{equation}
|F(Q^2)|^2=\frac{ ({\rm d}\sigma/{\rm d}Q^2)_{\rm data}}
{({\rm d}\sigma/{\rm d}Q^2)_{\rm MC}}F^2_{\rm MC }.
\end{equation}
The calculated cross section $({\rm d}\sigma/{\rm d}Q^2)_{\rm MC}$
has a model-dependent uncertainty due to the unknown dependence on the 
momentum transfer to the untagged electron. The difference between the cross
section values calculated with the two form-factor models described 
above is 4.6\% for both $\eta$ and
$\eta^\prime$. This difference is considered to be 
an estimate of the model uncertainty due to the unknown $q^2_2$ dependence.
The values of the form factors obtained, represented in the form
$\overline{Q^2}|F(\overline{Q^2})|$, are listed in Tables~\ref{tab30}
and \ref{tab31} and shown in Fig.~\ref{fig15}.
For the form factor we quote the combined error, obtained by adding
the statistical and $Q^2$-dependent systematic uncertainties
in quadrature. The $Q^2$-independent systematic error is
2.9\% for the $\eta$ and 3.5\% for the $\eta^\prime$ form factor. 

\section{Discussion and summary}
The comparison of our results on the form factors with
the most precise previous measurements~\cite{CLEO} is
shown in Fig.~\ref{fig15}. For the $\eta^\prime$ form factor 
our results are in good agreement with those reported
by the CLEO Collaboration~\cite{CLEO}. For the $\eta$
form factor the agreement is worse. In particular, the CLEO 
point at $Q^2\approx 7$ GeV$^2$ lies higher than
our measurements by about 3 standard deviations. 

\begin{figure*}
\includegraphics[width=.45\textwidth]{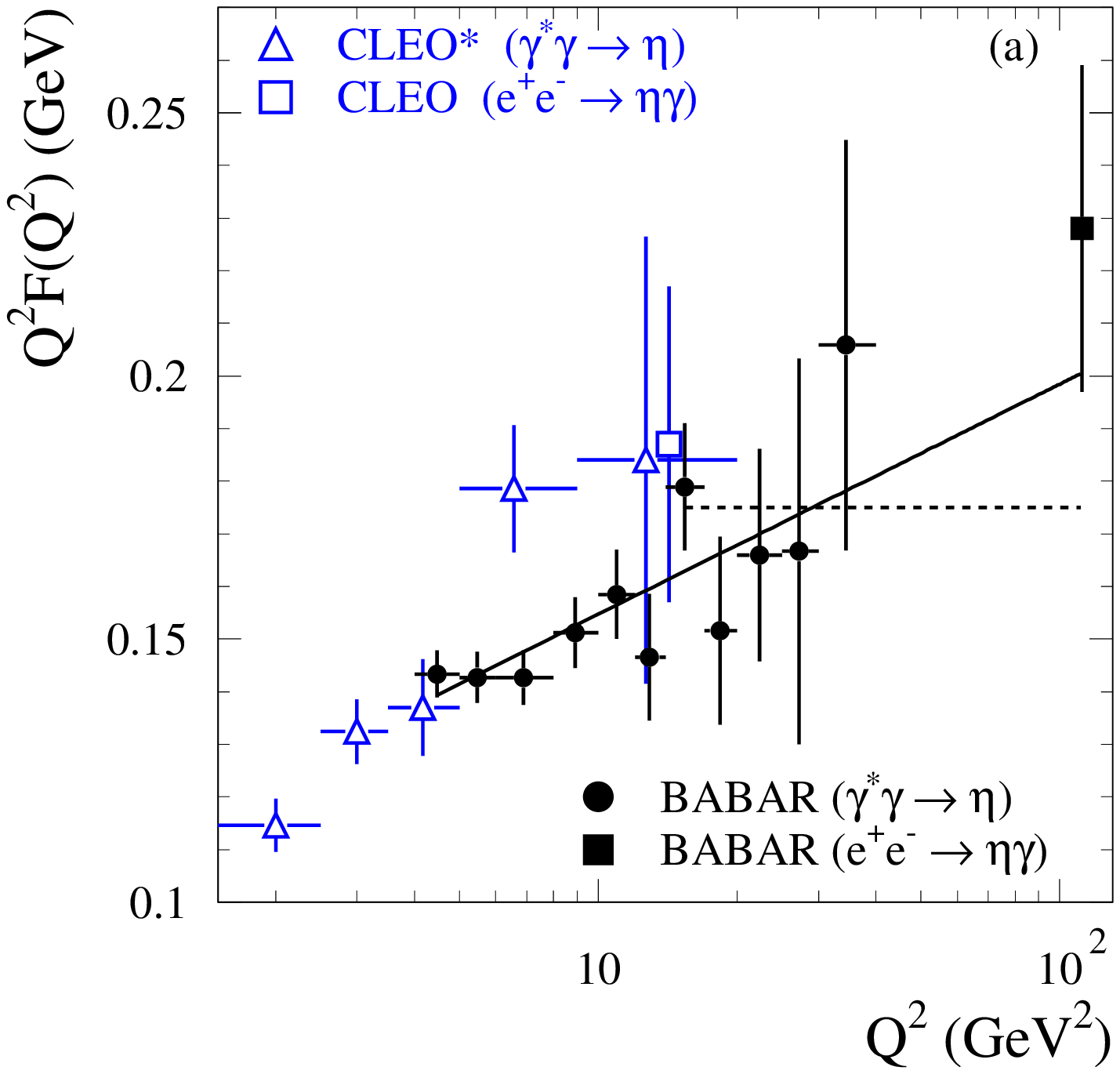}
\includegraphics[width=.45\textwidth]{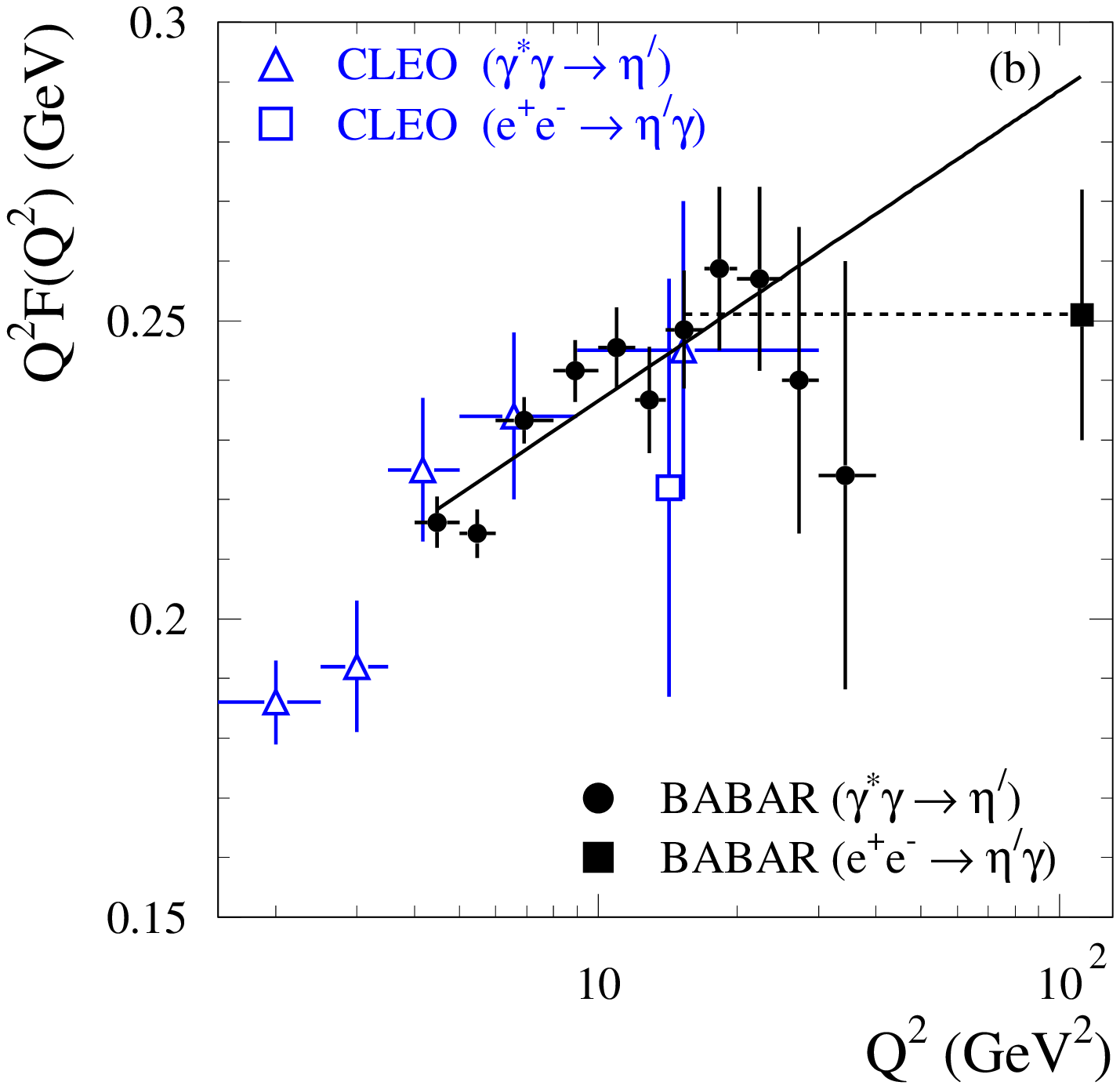}
\caption{The transition form factors multiplied by $Q^2$ for 
(a) $\gamma\gamma^\ast\to\eta$ and 
(b) $\gamma\gamma^\ast\to\eta^\prime$.
The solid line shows the result of the fit to
 \babar\ data by the function given by Eq.~(\ref{logr}).
The dashed lines indicate the average form factor
values over the data points with $Q^2 > 14$ GeV$^2$.
\label{fig16}}
\end{figure*}

The data for the $e^+e^- \to \eta^{(\prime)}\gamma$ reactions are used to 
determine the transition form factors in the time-like region $q^2 = s > 0$.
Since the time- and space-like form factors are expected to be similar
at high $Q^2$, in Fig.~\ref{fig16} we show the results of the high-$Q^2$ 
time-like measurements together with the space-like data.
The form factors at $Q^2=14.2$ GeV$^2$ are obtained from the values of the
$e^+e^- \to \eta^{(\prime)}\gamma$ cross sections measured by CLEO~\cite{etaff_4} near
the peak of the $\psi(3770)$ resonance. We calculate the form factor using the
formulas from Ref.~\cite{etaff_5} under the assumption that the
contributions of the $\psi(3770) \to \eta^{(\prime)}\gamma$ decays to the
$e^+e^- \to \eta^{(\prime)}\gamma$ cross sections are negligible.
It is seen that the measured time- and space-like form factors at 
$Q^2 \approx 14$ GeV$^2$ are in agreement both for $\eta$ and for $\eta^\prime$. 
The \babar\ measurements of the $e^+e^- \to \eta^{(\prime)}\gamma$ cross 
sections~\cite{etaff_5}
allow us to extend the $Q^2$ region for the $\eta$ and $\eta^\prime$ form factor
measurements up to 112 GeV$^2$.

In most models for the meson distribution amplitude $\phi_P(x)$ used for calculation 
of photon-meson transition form factors, the DA end-point behavior is
determined by the factor $x(1-x)$.
The form factors calculated with such conventional DAs are almost flat for $Q^2$ 
values greater than 15 GeV$^2$ (see, for example, the recent 
works~\cite{Mikhailov, Khodjamirian, Chernyak1} devoted
to the $\gamma\gamma^\ast \to \pi^0$ form factor).
Some of these models~\cite{Mikhailov} have difficulties in reproducing the $Q^2$
dependence of the $\gamma\gamma^\ast\to \pi^0$ form factor measured 
by \babar~\cite{pi0ff} in the $Q^2$ range from 4 to 40 GeV$^2$.
Alternatively, models with a flat DA or a DA that is finite at the end points
have been suggested~\cite{flat1,flat2,flat3},
which give a logarithmic rise of the product $Q^2 F(Q^2)$ with $Q^2$ and 
describe the  \babar\ data reasonably well.

The $Q^2$ dependencies of the products $Q^2 F(Q^2)$ for $\eta$ and $\eta^\prime$ 
are fit with the function 
\begin{equation}
Q^2F(Q^2)=b_l+a_l\ln Q^2({\rm GeV}^2).
\label{logr}
\end{equation}
The results of the fit are shown in Fig.~\ref{fig16}. For both $\eta$ and 
$\eta^\prime$ the quality of the fit is acceptable: 
$\chi^2/\nu$ is equal to 6.8/10 for $\eta$ and 15.9/10 for $\eta^\prime$,
where $\nu$ is the number of degrees of freedom.
The observed rise of the
form factors ($a_l\approx 0.20\pm0.05$ GeV) is about three times
weaker than the corresponding rise of the $\pi^0$ form factor~\cite{flat2}.

The dashed horizontal lines in Fig.~\ref{fig16} 
show the results of fits assuming $Q^2 F(Q^2)$ to be constant for
$14 < Q^2 < 112$ GeV$^2$. The average values of $Q^2 F(Q^2)$ in this range are
$0.175\pm0.008$ GeV for $\eta$ and $0.251\pm0.006$ GeV for $\eta^\prime$. 
The $\chi^2/\nu$ for the fits are 5.6/5 for the $\eta$ and 1.3/5 for the $\eta^\prime$. 
The preferred
description for the $\eta$ form factor is the logarithmic function 
of Eq.~(\ref{logr}), corresponding to the models with a finite DA at the end points.
The $\eta^\prime$ form factor is  better described   
by the model with a conventional DA, yielding a flat $Q^2 F(Q^2)$ for
$Q^2 > 15$ GeV$^2$. 

\begin{figure}
\includegraphics[width=.48\textwidth]{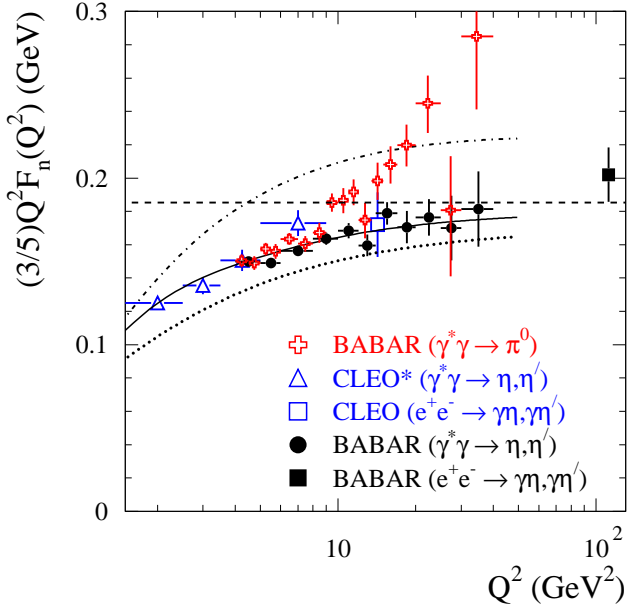}
\caption{The $\gamma\gamma^\ast\to |n\rangle$
transition form factor multiplied by $3Q^2/5$ in
comparison with the $\gamma\gamma^\ast\to \pi^0$
transition form factor~\cite{pi0ff}.
The dashed line indicates the asymptotic limit
for the $\pi^0$ form factor.
The dotted, dash-dotted, and solid curves
show predictions of Ref.~\cite{th6} for the asymptotic DA~\cite{ASY},
the Chernyak-Zhitnitsky $\pi^0$ DA~\cite{CZ},
and the $\pi^0$ DA from Ref.~\cite{BMS}, respectively.
\label{fig17}}
\end{figure}

To compare the measured values of the $\eta$ and $\eta^\prime$ form factors
with theoretical predictions and data for the $\pi^0$ form factor we use the 
description of
$\eta$-$\eta^\prime$ mixing in the quark flavor basis~\cite{mix1}:
\begin{eqnarray}
|n\rangle&=&\frac{1}{\sqrt{2}}(|\bar{u}u\rangle+|\bar{d}d\rangle),\;\;
|s\rangle=|\bar{s}s\rangle,\nonumber\\
|\eta\rangle& = &\cos{\phi}\, |n\rangle - \sin{\phi}\, |s\rangle,\nonumber\\
|\eta^\prime\rangle&  = &\sin{\phi}\, |n\rangle + \cos{\phi}\, |s\rangle,
\end{eqnarray}
where $\phi$ is the mixing angle.
The $\eta$ and $\eta^\prime$ transition form factors are related
to the form factors for the $|n\rangle$ and $|s\rangle$ states: 
\begin{equation}
F_\eta=\cos{\phi}\,F_n- \sin{\phi}\,F_s,\;\;
F_{\eta^\prime}=\sin{\phi}\,F_n+\cos{\phi}\,F_s, 
\end{equation}
which have asymptotic limits for $Q^2 \to \infty$~\cite{mix2} given by
\begin{equation}
Q^2F_s(Q^2)=\frac{2}{3}f_s,\;\;
Q^2F_n(Q^2)=\frac{5\sqrt{2}}{3}f_n,
\end{equation}
where $f_n$ and $f_s$ are the decay constants for
the $|n\rangle$ and $|s\rangle$ states, respectively. For the $\pi^0$ form factor,
the corresponding asymptotic value is $\sqrt{2}f_\pi$.  
The pion decay constant is determined from leptonic $\pi$ decays
to be $130.4\pm0.2$ MeV~\cite{pdg}. For the $|n\rangle$ and $|s\rangle$ states, 
we use the ``theoretical'' values from Ref.~\cite{mix1}: $f_n=f_\pi$ and
$f_s=\sqrt{2f_K^2-f_\pi^2}\approx1.36f_\pi$ ($f_K/f_\pi=1.193\pm0.006$~\cite{pdg}), 
which agree to within 10\% with the ``phenomenological''
values~\cite{mix1} extracted from the analysis of experimental data, 
for example, for the two-photon $\eta$ and $\eta^\prime$ decays. 
The currently accepted 
value of the mixing angle $\phi$ is about $41^\circ$~\cite{tomas}. 
Under the assumption that the $|n\rangle$ and $\pi^0$ distribution amplitudes
are similar to each other,
the only difference between the $|n\rangle$ and $\pi^0$ form factors is
a factor of $3/5$ that arises from the quark charges.
In Fig.~\ref{fig17} the form factor for the $|n\rangle$-state multiplied by
$3Q^2/5$ is compared with the measured $\gamma^\ast\gamma \to \pi^0$
form factor~\cite{pi0ff} and the results of the QCD calculations 
performed by A.~P.~Bakulev, S.~V.~Mikhailov and
N.~G.~Stefanis~\cite{th6} for the asymptotic DA~\cite{ASY},
the Chernyak-Zhitnitsky $\pi^0$ DA~\cite{CZ},
and the $\pi^0$ DA derived from QCD sum rules with non-local
condensates~\cite{BMS}. The horizontal dashed line indicates 
the asymptotic limit for the $\pi^0$ form factor. 
\begin{figure}
\includegraphics[width=.48\textwidth]{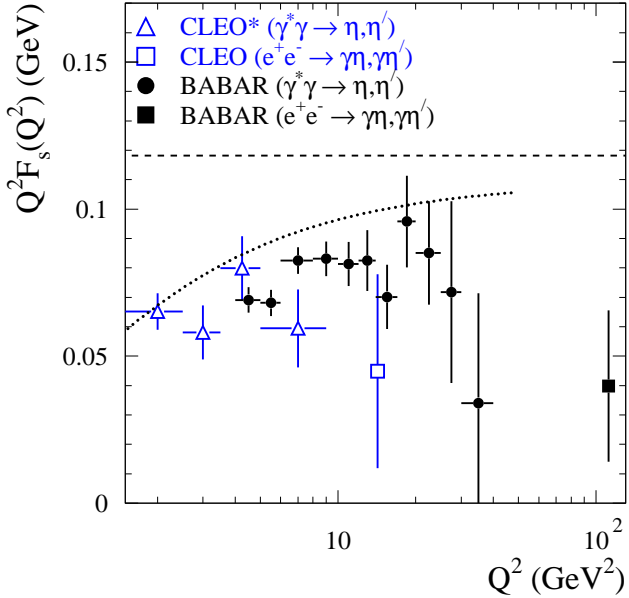}
\caption{The $\gamma\gamma^\ast\to |s\rangle$
transition form factor multiplied by $Q^2$.
The dashed line indicates the asymptotic limit
for the form factor. The dotted curve shows the prediction~\cite{th6}
for the asymptotic DA~\cite{ASY}.
\label{fig18}}
\end{figure}

The $Q^2$ dependencies of the measured $|n\rangle$ and $\pi^0$ 
form factors are significantly different. This indicates that the distribution 
amplitudes 
for the $|n\rangle$ and $\pi^0$ are significantly different as well.
The data for the $|n\rangle$ form factor are well
described by the model with DA from Ref.~\cite{BMS}, while 
the data for the $\pi^0$ form factor is reproduced by the models with a significantly 
wider DA~\cite{Khodjamirian, Chernyak1} or a flat DA~\cite{flat1,flat2,flat3}.

The form factor for the $|s\rangle$ state is shown in Fig.~\ref{fig18}. 
The dotted curve shows the QCD prediction~\cite{th6} for the asymptotic 
DA~\cite{ASY}, defined by multiplying the $\pi^0$ curve in Fig.~\ref{fig17}
by a factor of $(\sqrt{2}/3)f_s/f_\pi$.
The data lie systematically below this prediction. This may indicate,
in particular, that the distribution amplitude for the $|s\rangle$ state
is narrower than the asymptotic DA. However, due to the strong sensitivity of
the result for the the $|s\rangle$ state to mixing parameters, other 
interpretations are possible. For example, an admixture of the
two-gluon component in the $\eta^\prime$ meson~\cite{2g1,2g2,2g3,2g4} can
lead to a significant shift of the values of the $|s\rangle$ form factor.

In summary, we have studied the $e^+e^- \to e^+e^-\eta$ and
$e^+e^- \to e^+e^-\eta^\prime$ reactions and measured 
the differential cross sections (${\rm d}\sigma/{\rm d}Q^2$)
and the $\gamma^\ast\gamma\to \eta^{(\prime)}$ 
transition form factors $F(Q^2)$ in the momentum
transfer range from 4 to 40 GeV$^2$.
In general, our results are in reasonable agreement with the 
previous CLEO measurements~\cite{CLEO}.
We significantly improve the precision and extend the 
$Q^2$ region for form factor measurements.

\begin{acknowledgments}
We thank V. L. Chernyak for useful discussions. We are grateful for the 
extraordinary contributions of our \pep2\ colleagues in
achieving the excellent luminosity and machine conditions
that have made this work possible.
The success of this project also relies critically on the 
expertise and dedication of the computing organizations that 
support \babar.
The collaborating institutions wish to thank 
SLAC for its support and the kind hospitality extended to them. 
This work is supported by the
US Department of Energy
and National Science Foundation, the
Natural Sciences and Engineering Research Council (Canada),
the Commissariat \`a l'Energie Atomique and
Institut National de Physique Nucl\'eaire et de Physique des Particules
(France), the
Bundesministerium f\"ur Bildung und Forschung and
Deutsche Forschungsgemeinschaft
(Germany), the
Istituto Nazionale di Fisica Nucleare (Italy),
the Foundation for Fundamental Research on Matter (The Netherlands),
the Research Council of Norway, the
Ministry of Education and Science of the Russian Federation, 
Ministerio de Ciencia e Innovaci\'on (Spain), and the
Science and Technology Facilities Council (United Kingdom).
Individuals have received support from 
the Marie-Curie IEF program (European Union), the A. P. Sloan Foundation (USA) 
and the Binational Science Foundation (USA-Israel).
\end{acknowledgments}

\end{document}